\documentclass[cernpreprint, atlasdraft=false, coverpage=false, USenglish, texmf, orcidlogo]{atlasdoc}
\usepackage[multirow,longtable]{atlaspackage}
\usepackage{tabularx}
\usepackage{atlasbiblatex}

\usepackage[bsm,hepparticle]{atlasphysics}
\graphicspath{{logos/}{figures/}}

\usepackage{ANA-HMBS-2024-09-PAPER-defs}

\AtlasTitle{Search for charginos and neutralinos with $B-L$ $R$-parity violating decays in $\sqrt{s}=13$~TeV and $13.6$~TeV $pp$ collisions with the ATLAS detector}
\AtlasAbstract{%
A search is performed for the electroweak pair production of charginos and associated production of a chargino and neutralino, each of which decays through an $R$-parity-violating coupling into a lepton and a $W$, $Z$, or Higgs boson. This search targets the Higgs boson decay channel of the charginos and neutralinos, using events with three or more $b$-tagged jets and one or two electrons or muons.
The analyzed data corresponds to an integrated luminosity of 140~fb$^{-1}$ and 56~fb$^{-1}$ of proton--proton collision data produced by the Large
Hadron Collider at center-of-mass energies of $\sqrt{s}=13$~TeV and $\sqrt{s}=13.6$~TeV respectively, collected by the ATLAS
experiment between 2015 and 2023. The data are found to be consistent with predictions
from the Standard Model. The results are interpreted as limits at 95\% confidence level on
model-independent cross sections for processes beyond the Standard Model. Limits are also
set on the production of charginos and neutralinos for a Minimal Supersymmetric Standard
Model with an additional $B-L$ gauge symmetry that is spontaneously broken. Charginos and neutralinos with masses between
150~GeV and 1100~GeV are excluded at 95\% confidence level for a scenario in which they decay via Higgs bosons, assuming equal decay branching fractions to each lepton flavor. Additional limits on chargino and neutralino production are derived for flavor-specific decay scenarios where the charginos decay into a single lepton flavor.}

\AtlasRefCode{HMBS-2024-09}

\PreprintIdNumber{CERN-EP-2026-138}

\AtlasJournal{Phys.\ Rev.\ D.}
\AtlasJournalRef{Phys. Rev. D 114 (2026) 052007}
\AtlasDOI{10.1103/rp69-xgkd}

\AtlasCoverEgroupAnalysisTeam{atlas-hmbs-2024-09-analysis-team@cern.ch}

\hypersetup{pdftitle={ATLAS document},pdfauthor={The ATLAS Collaboration}}

\begin{document}

\maketitle

\tableofcontents

\section{Introduction}
\label{sec:intro}
%

% The next lines are included from the .//sections/introduction.tex input file
Supersymmetry (SUSY) is an extension of the Standard Model (SM)~\cite{Golfand:1971iw,Volkov:1973ix,Wess:1974tw,Wess:1974jb,Ferrara:1974pu,Salam:1974ig} that offers solutions to many open problems in particle physics, including the hierarchy problem~\cite{Sakai:1981gr,Dimopoulos:1981yj,Ibanez:1981yh,Dimopoulos:1981zb}. However, the full SUSY model space allows for baryon number (\(B\)) and lepton number (\(L\)) violation. While this can lead to rapid proton decay or large flavor-changing neutral currents in processes like \HepProcess{\Pgm\to\Pe\Pgg}, these processes can be prevented by requiring conservation of \Rparity~\cite{Farrar:1978xj}, a multiplicative quantum number defined as \(R=(-1)^{3(B-L)+2s}\), where \(s\) is the particle's spin. Enforcing the conservation of \Rparity (or equivalently, the quantity \(B-L\)) has additional phenomenological consequences for SUSY models. SM particles have \(R=+1\), whereas their SUSY partner particles (``superpartners'') have \(R=-1\). Therefore, in \Rparity conserving SUSY models, superpartners must be pair-produced at the Large Hadron Collider (LHC)~\cite{Evans:2008zzb}, and the lightest supersymmetric particle (LSP) must be stable.

\Rparity conservation, however, is an ad hoc solution, and may be too restrictive; there are phenomenologically viable \Rparity violating (RPV) SUSY theories. One such theory, known as the $B-L$ Minimal Supersymmetric Standard Model, introduces a \(\mathrm{U}(1)_{B-L}\) gauge symmetry in addition to the SM \(\mathrm{SU}(3)_C\times\mathrm{SU}(2)_L\times\mathrm{U}(1)_Y\) gauge group, along with right-handed sterile neutrinos~\cite{Marshall:2014kea,Deen:2016vyh,FileviezPerez:2015mlm,Ovrut:2015uea}. When the superpartners to the right-handed neutrino fields obtain a vacuum expectation value, the \(\mathrm{U}(1)_{B-L}\) symmetry is spontaneously broken in a way that violates only lepton number, while providing a mechanism to give the left-handed Standard Model neutrinos their small masses. RPV couplings allow direct decay of superpartners to SM particles, including the decay of the LSP, producing signatures that are forbidden in \Rparity conserving theories. These couplings remain weak, so the production mechanisms of superpartners are not significantly affected.

A wino LSP and next-to-lightest supersymmetric particle (NLSP), superpartners to the electroweak bosons whose primary component comes from the \PW boson superpartners, provide a unique search scenario for this model due to the possibility of a charged LSP\@. The charged component, the chargino \chinoonepm, and the neutral component, the neutralino \ninoone, are expected to be nearly mass-degenerate, with less than \qty{200}{\MeV} mass difference~\cite{Dumitru:2018jyb,Dumitru:2018nct}.
This small mass splitting suppresses \Rparity conserving decays of the NLSP to the LSP\@. The chargino undergoes prompt RPV decays into
a Higgs boson and a charged lepton (\(\chinoonepm\rightarrow\Ph\Pl^\pm\)),
to a \PZ boson and a charged lepton (\(\chinoonepm\rightarrow\PZ\Pl^\pm\)),
or to a \PW boson and a neutrino (\(\chinoonepm\rightarrow\PW^\pm\Pgn\)).
Similarly, the neutralino undergoes prompt decays into
a Higgs boson and a neutrino (\(\ninoone\rightarrow\Ph\Pgn\)),
to a \PZ boson and a neutrino (\(\ninoone\rightarrow\PZ\Pgn\)),
or to a \PW boson and a charged lepton (\(\ninoone\rightarrow\PW^\pm\Pl^\mp\)).
While the proportion of these decays depends on parameters of the SUSY theory, the branching fractions for decays for \(\chinoonepm\rightarrow\Ph\Pl^\pm\) and \(\ninoone\rightarrow\Ph\Pgn\) are expected to be above 80\% for most scenarios~\cite{Dumitru:2018nct} and are therefore targeted in this search. The relative branching fractions of these decays for different lepton flavors depends on the neutrino mass hierarchy~\cite{Dumitru:2018nct}.

\begin{figure}[htbp]
\centering
\subfloat[]{
\includegraphics[width=0.3\linewidth]{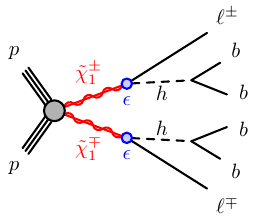}
\label{fig:feyndiag-C1C1}
}
\subfloat[]{
\includegraphics[width=0.3\linewidth]{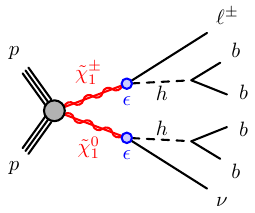}
\label{fig:feyndiag-C1N1}
}
\caption{%
Representative diagrams of the signal model processes considered in this analysis with \protect\subref{fig:feyndiag-C1C1} the chargino--chargino process (\(\chinoonepm\chinoonemp\)) and \protect\subref{fig:feyndiag-C1N1} the chargino--neutralino process (\(\chinoonepm\ninoone\)). The RPV effective vertices are marked with the bilinear RPV coupling \(\epsilon\).
}
\label{fig:feyndiag-C1X1}
\end{figure}

This paper presents a search for chargino-chargino pair production (\(\chinoonepm\chinoonemp\)) and chargino-neutralino production (\(\chinoonepm\ninoone\)) with RPV $\chinoonepm \to h \ell^\pm$ and $\ninoone \to h \nu$ decays. The data was collected with the ATLAS detector in proton--proton ($pp$) collisions at the LHC and comprises \qty{140}{\ifb} at $\sqrt{s}=13$~\TeV taken from 2015--2018 (\RunTwo) and \qty{56}{\ifb} at $\sqrt{s}=13.6$~\TeV taken from 2022--2023 (partial \RunThr). This analysis only considers the dominant decay of the Higgs boson into \bquarks ($\mathcal{B}(h\rightarrow b\bar{b})=58\%$). Higgs bosons are reconstructed from pairs of jets, and the selection requires at least three jets to be identified as originating from $b$-hadrons ($b$-tagged jets). Diagrams for the signal model are shown in Figure~\ref{fig:feyndiag-C1X1}. In the case of \(\chinoonepm\chinoonemp\) production, final states should contain two leptons with opposite-sign electric charge, while for \(\chinoonepm\ninoone\) production, final states should contain one lepton and missing transverse momentum ($\vec{p}_\text{T}^\text{miss}$, with magnitude \met), due to the undetected neutrino. While $\tau$-leptons are not explicitly targeted by this search, they are included in the signal sample generation and contribute to the sensitivity through their leptonic decays. The search strategy aims at reconstructing the mass of the chargino \mchinoone from Higgs--lepton resonances and the transverse mass of the neutralino \mTninoone from a Higgs boson and the $\vec{p}_\text{T}^\text{miss}$.

A previous ATLAS search for electroweakino pair production with RPV decays targeted chargino decays into $\PZ\ell$ with subsequent $Z\rightarrow \ell \ell$ by searching for a trilepton resonance using the full \RunTwo dataset~\cite{SUSY-2018-36}.
The search described in this paper provides complementary sensitivity by targeting chargino and neutralino decays into Higgs bosons and leptons, making use of different regions of phase space.

A dedicated series of \RunTwo searches for RPV \(B-L\) scenarios has also been carried out considering the superpartner of the top quark to be the LSP~\cite{SUSY-2016-29,SUSY-2018-37}. These were optimized for the pair production of top squarks with RPV decays into a bottom quark and charged lepton. The final state signature is a pair of resonances formed from a $b$-jet and a lepton, requiring $\ge1$ $b$-jets in the signal region. %

This analysis benefits from developments in the capability of identifying jets originating from $b$-hadrons, namely the GN2 algorithm~\cite{FTAG-2023-05}, which allows significantly better background reduction as compared with the previous algorithm. The analysis also benefits from strong background rejection due to the requirement of reconstructing two Higgs boson candidates. The sensitivity in the \(\chinoonepm\ninoone\) channel is driven by low background due to tight selections, while the sensitivity in the \(\chinoonepm\chinoonemp\) channel is due to good mass resolution for the chargino resonance from the reconstruction algorithms.

This paper is organized as follows. The ATLAS detector is described in Section~\ref{sec:detector}.
The data and simulated samples used are described in Section~\ref{sec:data_mc_samples}.
Analysis objects are defined in Section~\ref{sec:object_reco}.
The definition of analysis regions and the background estimation strategy are described in Section~\ref{sec:event_sel} and Section~\ref{sec:bkg} respectively.
Systematic uncertainties are discussed in Section~\ref{sec:systematics}, and analysis results are shown in Section~\ref{sec:result}.
Concluding remarks are made in Section~\ref{sec:conclusion}.

% End of text imported from the .//sections/introduction.tex input file

%
\section{ATLAS detector}
\label{sec:detector}
%

% The next lines are included from the .//sections/atlas.tex input file
%
%

%
\newcommand{\AtlasCoordFootnote}{%
ATLAS uses a right-handed coordinate system with its origin at the nominal interaction point (IP)
in the center of the detector and the \(z\)-axis along the beam pipe.
The \(x\)-axis points from the IP to the center of the LHC ring,
and the \(y\)-axis points upwards.
Polar coordinates \((r,\phi)\) are used in the transverse plane,
\(\phi\) being the azimuthal angle around the \(z\)-axis.
The pseudorapidity is defined in terms of the polar angle \(\theta\) as \(\eta = -\ln \tan(\theta/2)\) and is equal to the rapidity
$ y = \frac{1}{2} \ln \left( \frac{E + p_z}{E - p_z} \right) $ in the relativistic limit.
Angular distance is measured in units of \(\Delta R \equiv \sqrt{(\Delta y)^{2} + (\Delta\phi)^{2}}\).}

The ATLAS experiment~\cite{PERF-2007-01,GENR-2019-02} at the LHC is a multipurpose particle detector
with a forward--backward symmetric cylindrical geometry and a near \(4\pi\) coverage in
solid angle.\footnote{\AtlasCoordFootnote}
It consists of an inner tracking detector (ID) surrounded by a thin superconducting solenoid
providing a \qty{2}{\tesla} axial magnetic field, electromagnetic and hadronic calorimeters, and a muon spectrometer (MS).
The inner tracking detector covers the pseudorapidity range \(|\eta| < 2.5\).
It consists of silicon pixel (including the insertable B-layer), silicon microstrip, and transition radiation tracking detectors.
Lead/liquid-argon (LAr) sampling calorimeters provide electromagnetic (EM) energy measurements
with high granularity within the region \(|\eta|< 3.2\).
A steel/scintillator-tile hadronic calorimeter covers the central pseudorapidity range (\(|\eta| < 1.7\)).
The endcap and forward regions are instrumented with LAr calorimeters
for EM and hadronic energy measurements up to \(|\eta| = 4.9\).
The muon spectrometer surrounds the calorimeters and is based on
three large superconducting air-core toroidal magnets with eight coils each.
The field integral of the toroids ranges between \num{2.0} and \qty{6.0}{\tesla\metre}
across most of the detector.
The muon spectrometer includes a system of precision tracking chambers up to \(|\eta| = 2.7\) and fast detectors for triggering up to \(|\eta| = 2.4\).
The luminosity is measured mainly by the LUCID--2 detector that is located close to the beampipe.
A two-level trigger system was used to select events~\cite{TRIG-2016-01,TRIG-2022-01}.
The first-level trigger is implemented in hardware and used a subset of the detector information
to accept events at a rate close to \qty{100}{\kHz}.
This is followed by a software-based trigger that
reduced the accepted rate of complete events to \qty{1.25}{\kHz} and \qty{3}{\kHz} on average in Run~2 and Run~3, respectively,
depending on the data-taking conditions.
A software suite~\cite{SOFT-2022-02} is used in data simulation, in the reconstruction
and analysis of real and simulated data, in detector operations, and in the trigger and data acquisition
systems of the experiment.

% End of text imported from the .//sections/atlas.tex input file

%
\section{Data and simulated samples}
\label{sec:data_mc_samples}

% The next lines are included from the .//sections/data_mc.tex input file
This analysis is performed using $pp$ collision data collected by the ATLAS experiment during \RunTwo of the LHC from 2015 to 2018 and \RunThr of the LHC from 2022 to 2023. The LHC collided protons at a center-of-mass energy of $\sqrt{s}=13$~\TeV during \RunTwo and $\sqrt{s}=13.6$~\TeV during \RunThr. The data sample corresponds to a total integrated luminosity of 196 \ifb, with 140 \ifb from \RunTwo and 56 \ifb from \RunThr. The uncertainty in the 2015--2018 (2022--2023) integrated luminosity is 0.83\%~\cite{DAPR-2021-01} (2.0\%~\cite{ATL-DAPR-PUB-2024-001}), obtained using the LUCID-2 detector~\cite{LUCID2} for the primary luminosity measurements, complemented by measurements using the inner detector and calorimeters. Data quality requirements are applied to ensure that only data with stable LHC beams and operational subdetectors are used~\cite{DAPR-2018-01}. Data were collected using triggers requiring the presence of an electron or a muon~\cite{TRIG-2018-05, TRIG-2018-01, TRIG-2022-01}. The triggers have varying requirements on the identification working point, isolation, and transverse momentum ($\pT$) thresholds of the leptons.

Monte Carlo (MC) simulations are used to model the signals and SM background processes for this analysis. All samples were produced with the ATLAS simulation framework~\cite{SOFT-2010-01, SOFT-2022-02} based on \GEANT~\cite{Agostinelli:2002hh}. The signal samples simulate $\chinoonepm\chinoonemp$ and $\chinoonepm\ninoone$ production at leading order (LO) with up to two additional partons in the matrix element. In the benchmark model considered, the masses of the chargino and neutralino are set to be equal. The production cross sections are computed at next-to-leading-order (NLO) plus next-to-leading-logarithm (NLL) precision in a limit of mass-degenerate wino $\chinoonepm$ and $\ninoone$, with all the other sparticles assumed to be heavy and decoupled~\cite{Fuks:2012qx,Fuks:2013vua,Fiaschi:2023tkq}. These vary from about 2612~fb (5181~fb) for $\chinoonepm\chinoonemp$ ($\chinoonepm\ninoone$) production at a chargino and neutralino mass of $\qty{150}{\gev}$ to about 0.62~fb (1.34~fb) for $\chinoonepm\chinoonemp$ ($\chinoonepm\ninoone$) production at $\qty{1000}{\gev}$ for \RunTwo, with slightly higher cross sections for \RunThr. The decay branching ratios are set to $\mathcal{B}(\chinoonepm\rightarrow h\ell^\pm)=100\%$ and $\mathcal{B}(\ninoone\rightarrow h\nu)=100\%$, with equal branching fractions to each lepton generation.
Samples were generated with \MGNLO[3.5.3]~\cite{Alwall:2014hca}
with the \NNPDF[3.0nlo]~\cite{Ball:2014uwa} parton distribution function (PDF). The parton shower, hadronization, and underlying event were modeled using \PYTHIA[8.310]~\cite{Bierlich:2022pfr} with the A14 set of tuned parameters (tune)~\cite{ATL-PHYS-PUB-2014-021}.
The MC setup for the signal, along with the SM background processes, is shown in Table~\ref{tab:MC}.

\begin{table}[ht]
\footnotesize
\centering
\caption{
Details of the MC simulation for each physics process, including the calculations used for normalization,
the generator used for the parton shower (PS) and hadronization, the PS parameter tunes, and the order in \alphas of the production cross-section calculations. NLO stands for next-to-leading-order, NNLO stands for next-to-next-to-leading-order, NLL stands for next-to-leading-logarithm, and NNLL stands for next-to-next-to-leading-logarithm.
The default table values are for \RunTwo, with \RunThr in parentheses if they differ.
}
\label{tab:MC}

\resizebox{\columnwidth}{!}{%
\begin{tabular}{lcccc}
\toprule
\multirow{2}{*}{Process} & \multirow{2}{*}{Event generator}    & PS and        & \multirow{2}{*}{PS tune} & \multirow{2}{*}{Cross section accuracy} \\
&                                     & hadronization &                          & \\
\hline
\hline

$\chinoonepm\chinoonemp$ and $\chinoonepm\ninoone$
& \MGNLO[3.5.3]~\cite{Alwall:2014hca}
& \PYTHIA[8.310]~\cite{Bierlich:2022pfr}
& A14~\cite{ATL-PHYS-PUB-2014-021}
& \scriptsize{NLO+NLL} \\ %

\ttbb
& \POWHEGBOXRES~\cite{Jezo:2018yaf} and \OPENLOOPS~\cite{Buccioni:2019sur,Cascioli:2011va,Denner:2016kdg}
& \PYTHIA[8.309(8.312)]
& A14
& \scriptsize{NLO} \\ %

\ttbar
& \POWHEGBOX[v2]~\cite{Frixione:2007nw,Nason:2004rx,Frixione:2007vw,Alioli:2010xd}
& \PYTHIA[8.230(8.308)~\cite{Sjostrand:2014zea}]
& A14
& \scriptsize{NNLO+NNLL}~\cite{Beneke:2011mq,Cacciari:2011hy,Baernreuther:2012ws,Czakon:2012zr,Czakon:2012pz,Czakon:2013goa,Czakon:2011xx} \\ %

\tth
& \POWHEGBOX[v2]
& \PYTHIA[8.230(8.308)]
& A14
& \scriptsize{NLO}~\cite{deFlorian:2016spz} \\ %

\ttW
& \SHERPA[2.2.14]~\cite{Bothmann:2019yzt}
& \SHERPA[2.2.14]
& Default
& \scriptsize{NLO}~\cite{deFlorian:2016spz} \\

\ttZ
& \MGNLO[2.8.1(3.5.1)]
& \PYTHIA[8.210(8.309)]
& A14
& \scriptsize{NLO}~\cite{deFlorian:2016spz} \\

Single-top (\(s\)/\(t\)-channel)
& \POWHEGBOX[v2]
& \PYTHIA[8.230(8.308)]
& A14
& \scriptsize{NNLO}~\cite{ATL-PHYS-PUB-2025-035} \\ %

Single-top (\(\PW\Pqt\))
& \POWHEGBOX[v2]
& \PYTHIA[8.309(8.308)]
& A14
& \scriptsize{NLO+NNLL}~\cite{Kidonakis:2010ux,Kidonakis:2013zqa} \\ %

Diboson
& \SHERPA[2.2]
& \SHERPA[2.2]
& Default
& \scriptsize{NLO} \\

$V$+jets
& \SHERPA[2.2.11(2.2.14)]
& \SHERPA[2.2.11(2.2.14)]
& Default
& \scriptsize{NNLO}~\cite{Anastasiou:2003ds} \\

\bottomrule
\end{tabular}}
\end{table}

The dominant background in this analysis is from top-quark pair production (\ttbar). The production of \ttbar events was modeled using the
\POWHEGBOX[v2]~\cite{Frixione:2007nw,Nason:2004rx,Frixione:2007vw,Alioli:2010xd}
generator at NLO with the \NNPDF[3.0nlo] PDF set
and the \hdamp parameter\footnote{The
\hdamp parameter is a resummation damping factor and one of the
parameters that controls the matching of \POWHEG matrix elements to
the parton shower. It thus effectively regulates the
high-\pT radiation against which the \ttbar system recoils.} set
to 1.5\,\mtop~\cite{ATL-PHYS-PUB-2016-020}.  The events were interfaced
to \PYTHIA[8.230(8.308)]~\cite{Sjostrand:2014zea} for \RunTwo (\RunThr) to model the parton shower,
hadronization, and underlying event, with parameters set according
to the A14 tune and using the \NNPDF[2.3lo]
set of PDFs~\cite{Ball:2012cx}.

These \ttbar events are separated by particle flavor into \ttbar with additional light-flavor jets (\ttlight), \cjets (\ttc), or \bjets (\ttb). This classification uses generator-level jets, which are reconstructed from stable particles using the \antikt algorithm~\cite{Cacciari:2008gp, Fastjet} with a radius parameter of \(R=0.4\). The filtering classifies the flavor of generator-level jets by using a \(\Delta R\) matching with heavy-flavor hadrons (excluding hadrons with the top quark or $W$ boson as parent particles). Generator-level jets with \(\pt > \qty{15}{\GeV}\) and \(\abseta < 2.5\) are labeled $b$-flavored if they are matched within \(\Delta R = 0.4\) to a $b$-hadron with \(\pt > \qty{5}{\GeV}\), and events containing at least one such jet are labeled \ttb events. Jets that are not matched to a $b$-hadron but are matched to a $c$-hadron with \(\pt > \qty{5}{\GeV}\) within \(\Delta R = 0.4\) are labeled $c$-flavored, and events containing at least one such jet and no $b$-flavored jets are labeled \ttc events. All \ttbar events that do not contain a $b$-flavored or $c$-flavored jet are classified as \ttlight events. The \ttb events are removed from the inclusive \ttbar sample and are instead modeled with dedicated \ttbb samples. The \ttc and \ttlight events are removed from the \ttbb samples. These \ttbb events were produced using the \POWHEGBOXRES~\cite{Jezo:2018yaf} generator at NLO
and \OPENLOOPS~\cite{Buccioni:2019sur,Cascioli:2011va,Denner:2016kdg},
and the \NNPDF[3.1nlo]~\cite{NNPDF:2017mvq} PDF set.
The events were interfaced to \PYTHIA[8.309(8.312)] in \RunTwo (\RunThr)
using the A14 set of tuned parameters
and using the \NNPDF[2.3lo] set of PDFs.
The four-flavor scheme was used with the \bquark mass set to \qty{4.95}{\GeV}.

The production of \ttH events was modeled using the
\POWHEGBOX[v2]~\cite{Hartanto:2015uka}
generator at NLO with the \NNPDF[3.0nlo] PDF set.
The production of \ttZ events was modeled using the
\MGNLO[ 2.8.1 (3.5.1)] generator at NLO with the
\NNPDF[3.0nlo] PDF in \RunTwo (\RunThr).
The production of \ttW events was modeled using \SHERPA[2.2.14]~\cite{Bothmann:2019yzt} with up to one additional parton at NLO and two additional partons at LO and the \NNPDF[3.0nnlo]~\cite{Ball:2014uwa} PDF set. Events from the \ttH, \ttZ, and \ttW processes are collectively labeled \ttX.

Single-top \(s\)- and \(t\)-channel production were modeled using the
\POWHEGBOX[v2]~\cite{Alioli:2010xd}
generator at NLO in QCD in the five-flavor scheme with the \NNPDF[3.0nlo] PDF set, while \(\PW\Pqt\) samples were generated using \POWHEGBOX[v2] at NLO interfaced with \PYTHIA[8.309] (\PYTHIA[8.308]) for \RunTwo (\RunThr). The \(\PW\Pqt\) samples used the diagram removal scheme~\cite{Frixione:2008yi}.
Diboson production was modeled using the \SHERPA[2.2.14] or 2.2.16 generator depending on the process, including off-shell effects and Higgs boson contributions, where appropriate. Fully leptonic final states and semileptonic final states were generated using matrix elements at NLO accuracy in QCD for up to one additional parton and at LO accuracy for up to three additional parton emissions. Samples for $gg\rightarrow VV$ were generated for \RunTwo using \SHERPA[2.2.2] with LO-accurate matrix elements for up to one additional parton emission for both the cases of fully leptonic and semileptonic final states. The \NNPDF[3.0nnlo] PDF set was used for all diboson samples.
The production of \Vjets was modeled using
\SHERPA[2.2.14] in \RunThr as well as in \RunTwo \HepProcess{\PZ\to\Pgt\Pgt}, and \SHERPA[2.2.11] for the remainder of \RunTwo processes, with up to two additional partons at NLO and up to five additional partons at LO\@.
The virtual QCD corrections were provided by the
\OPENLOOPS library.
The \NNPDF[3.0nlo] PDF set was used.

For all samples using \SHERPA, the matrix elements were merged with the parton shower using the \MEPSatNLO prescription~\cite{Hoeche:2011fd,Hoeche:2012yf,Catani:2001cc,Hoeche:2009rj}. For all samples using \PYTHIA for the parton shower model, the decays of bottom and charm hadrons were simulated using \EVTGEN~\cite{Lange:2001uf}. The effect of multiple interactions in the same and neighboring bunch crossings (\pileup) was modeled by overlaying~\cite{SIMU-2020-01} the simulated hard-scattering event
with inelastic \(pp\) events generated from a mix of
\EPOS[2.0.1.4]~\cite{Werner:2005jf} and \PYTHIA[8.308].
The \EPOS events were generated with the \EPOS LHC tune~\cite{Pierog:2013ria}
and the \PYTHIA events with the A3 tune~\cite{ATL-PHYS-PUB-2016-017}
and the \NNPDF[2.3lo] set of PDFs.
\PYTHIA pileup events include either a high-\pt jet,
a prompt photon, or a lepton from a $b$-hadron decay,
while \EPOS was filtered to simulate all remaining pileup events in the overlay sample.
The individual simulations were first reweighted to ensure a smooth connection across jet \pt then
the combination reweighted to match the distribution of the actual number of interactions per bunch crossing (\(\mu\)) measured in data.

% End of text imported from the .//sections/data_mc.tex input file

%

%
\section{Object reconstruction}
\label{sec:object_reco}

% The next lines are included from the .//sections/object.tex input file
%

Charged-particle tracks are required to have $\pt > \SI{0.5}{\gev}$~\cite{PERF-2015-08, IDTR-2022-04, IDTR-2021-01}. Primary vertex candidates are reconstructed from at least two charged-particle tracks~\cite{ATL-PHYS-PUB-2015-026}. To identify the hard-scattering interaction, the event's primary vertex is chosen as the vertex with the largest sum of the squared track \pt ($\sum p_\text{T,track}^2$).

Electrons, muons, and jets are defined for this analysis at two levels: ``baseline'' and ``signal-quality''. Baseline objects are used as inputs to the $\vec{p}_\text{T}^\text{miss}$ calculation and overlap removal, while signal-quality objects are used for event selections. A summary of the $\pt$ and $\eta$ requirements on signal-quality objects is shown in Table~\ref{tab:objects}. Electrons are reconstructed as described in Ref.~\cite{EGAM-2018-01} and are calibrated based on the procedure and results described in Refs.~\cite{EGAM-2018-01, EGAM-2021-02}. For \RunThr, there are additional corrections on the \textit{in situ} scale and resolution corrections that account for small differences in the ATLAS reconstruction software, increased uncertainties to reflect the \RunThr pileup conditions, and a change in optimal filtering coefficients for the LAr calorimeter readout. %
Baseline electrons are required to satisfy the \textit{Loose} likelihood identification criteria with at least one B-layer hit~\cite{EGAM-2021-01} and have \(\pt > \qty{10}{\GeV}\), \(\abseta < 2.47\), and a selection of \(|\zzsth| < \qty{0.5}{\milli\meter}\) on the longitudinal impact parameter \(z_0\). Baseline electrons that survive the overlap removal procedure described at the end of this section, satisfy the \textit{Loose} (\textit{Tight}) isolation requirements~\cite{EGAM-2021-01} in \RunTwo (\RunThr), satisfy the \textit{Tight} identification criteria, have \(\pt > \qty{40}{\GeV}\), and have a transverse impact parameter significance \(|d_0|/\sigma(d_0) < 5\) are labeled as signal-quality electrons. The large \pt threshold is used to suppress jets and photons misidentified as electrons.

Muons are identified using tracks in the ID and MS, and are calibrated to match data~\cite{MUON-2022-01, PERF-2015-10}. Baseline muons are required to satisfy the \textit{Medium} identification requirement~\cite{MUON-2018-03} and have \(\pt > \qty{10}{\GeV}\), \(\abseta < 2.7\), and a selection of \(|\zzsth| < \qty{0.5}{\milli\meter}\). For \RunThr, baseline muons are also required to have \(\abseta < 2.5\). Other selections and requirements match those used in \RunTwo. Muons with transverse impact parameter $|d_0|\ge\qty{0.2}{\milli\meter}$ or longitudinal impact parameter $|z_0|\ge\qty{1}{\milli\meter}$ are removed to reject muons originating from cosmic-ray interactions, and events containing baseline muons flagged as bad due to misalignments between the inner detector and muon spectrometer are vetoed. Baseline muons that survive overlap removal, satisfy the \textit{Loose} isolation working point, satisfy the \textit{HighPt} identification criteria~\cite{MUON-2018-03}, have \(\pt >\qty{40}{\GeV}\), and have transverse impact parameter significance \(|d_0|/\sigma(d_0) < 3\) are labeled signal-quality muons. Events with signal-quality muons flagged as bad are again removed, as the criteria are different for the \textit{HighPt} working point. The large \pt threshold is used to suppress objects misidentified as muons. %

Jets are reconstructed using the \antikt algorithm~\cite{Cacciari:2008gp, Fastjet} with a radius parameter of \(R=0.4\). This uses particle-flow objects~\cite{PERF-2015-09} as inputs, which are formed using calorimeter energy clusters~\cite{PERF-2014-07} and ID tracks.
Calibrations are applied for the jet energy scale and jet energy resolution, and include components derived both from simulation and \textit{in situ} measurements~\cite{JETM-2018-05}. Events containing jets from non-collision backgrounds are vetoed~\cite{ATLAS-CONF-2015-029}.
Baseline jets are required to have \(\pt > \qty{20}{\GeV}\) and \(\abseta < 2.8\). Baseline jets are used in the overlap removal described below. Baseline jets that survive overlap removal are subjected to jet vertex tagger (JVT) requirements to suppress pileup, with remaining jets labeled signal-quality jets. Jets with \(\pt < 60\) \(\gev\) and \(|\eta| < 2.5\) are subjected to the \textit{FixedEffPt} JVT working point~\cite{PERF-2014-03} while jets with \(\pt < \qty{60}{\GeV}\) and \(|\eta| > 2.5\) use the \textit{Loose} forward jet vertex tagger (fJVT) working point~\cite{PERF-2016-06}.

Jets containing $b$-hadrons (\bjets) are identified using the GN2 tagger at a working point corresponding to 77\% efficiency as measured in \ttbar events~\cite{FTAG-2023-05}. The tagger uses a transformer-based model utilizing information about secondary vertices and other jet and track properties to identify \bjets.%

Baseline photons are reconstructed as described in Ref.~\cite{EGAM-2018-01}. They are required to satisfy the \textit{Tight} identification working point~\cite{EGAM-2021-01,EGAM-2018-01} and have \(\pt > \qty{25}{\GeV}\) and \(\abseta < 2.37\). The \textit{Tight} identification working point removes photons in the crack region of the detector, \(1.37<\abseta<1.52\). Baseline photons are used in the $\vec{p}_\text{T}^\text{miss}$ calculation, but not in the overlap removal procedure described below nor in any analysis selections.

The missing transverse momentum $\vec{p}_\text{T}^\text{miss}$, with magnitude \met, is defined as the negative vector sum of the \pt of all selected and calibrated baseline physics objects in the event, with an extra term added to account for soft energy in the event that is not associated with any of the selected objects~\cite{JETM-2020-03}. This soft term is calculated from inner detector tracks matched to the primary vertex to make it more resilient to pileup contamination. Jets that fail to meet the JVT requirement are not considered in the $\vec{p}_\text{T}^\text{miss}$ calculation, while jets that fail to meet the fJVT requirement are considered.

A procedure is applied to remove overlaps between baseline electrons, muons, and jets. First, any muon that is identified using the calorimeter and shares an ID track with an electron is rejected. Next, any electron sharing an ID track with a remaining muon is removed. If a jet is found to be within \(\Delta R = 0.2\) of a remaining electron, the jet is removed. If an electron is found to be within \(\Delta R = 0.4\) from a remaining jet, the electron is removed. Next, any jet with an associated muon ID track or a muon within \(\Delta R = 0.2\) of its axis is removed if the jet has less than three tracks. Lastly, any muon is removed if it is found to be within \(\Delta R  = 0.4\) from any remaining jet.

\begin{table}[t]
\centering
\caption{Summary of the $\pt$ and $\eta$ requirements for signal-quality objects. Signal-quality electrons and jets use the same requirements in \RunTwo and \RunThr.}
\label{tab:objects}
\begin{tabular}{ccc}
\toprule
Signal-quality Object  &  \pt[\GeV] & $|\eta|$ \\ \midrule
Electrons & \(> 40\) &  \(< 2.47\) \\
Muons (\RunTwo) & \(> 40\) &  \(< 2.7\) \\
Muons (\RunThr) & \(> 40\) &  \(< 2.5\) \\
Jets & \(> 20\) & \(< 2.8\) \\ \bottomrule
\end{tabular}
\end{table}

% End of text imported from the .//sections/object.tex input file

%
\section{Event selection}
\label{sec:event_sel}

% The next lines are included from the .//sections/event_selection.tex input file
%

%

The event selection for $\chinoonepm\chinoonemp\rightarrow h\ell^\pm h\ell^\mp \rightarrow b\bar{b}\ell^\pm b\bar{b}\ell^\mp$ and $\chinoonepm\ninoone\rightarrow h\ell h\nu \rightarrow b\bar{b}\ell b\bar{b}\nu$ requires events to have at least four jets, at least two of which must be identified as \bjets, and have at least one electron or muon. If there are two or more electrons or muons in an event, the two leading leptons are required to have opposite charge. Reconstructed signal-quality leptons, defined in Section~\ref{sec:object_reco}, have tighter identification working point requirements than applied at online level to ensure high trigger efficiency for selected events. Events are required to have at least one reconstructed signal-quality lepton that activated one of the triggers and has an offline $\pT$ higher than a trigger-dependent offline threshold. The lowest \pt thresholds are 27~\gev (27.3~\gev) for electrons (muons) in 2015--2018, while for 2022--2023 they are 27~\gev (25.2~\gev).

\subsection{Analysis variables}

Higgs boson candidates are reconstructed using four jets. If there are four or more $\bjets$ in an event, the four with the highest \pT are used. If there are fewer than four \bjets, all of the \bjets in the event are used, with the highest \pT non-$b$-tagged jets used for the remaining Higgs boson candidate jets. The jets are paired to form two Higgs boson candidates by looking for pairs of jets with a small opening angle \(\Delta R_{jj}(h)\), where \(\Delta R_{jj}(h)\) is the angular distance between the two jets from the decay of the Higgs boson \Ph and is calculated for each possible pairing. The jet pairing that minimizes \(\max(\Delta R_{jj}(h_1),\Delta R_{jj}(h_2))\) is chosen. Studies on generator-level signal MC simulations show that this algorithm makes the correct assignment around 40\% of the time for low chargino masses, rising to 95\% of the time for high chargino masses. The highest \pT Higgs boson candidate is labeled \(h_1\) and the other is labeled \(h_2\). The reconstructed masses of the Higgs boson candidates are referred to as \mhone and \mhtwo, respectively.

Higgs boson candidates, electrons, muons, and $\vec{p}_\text{T}^\text{miss}$ are used to reconstruct chargino and neutralino candidates. First, the four-momentum of each Higgs boson candidate is rescaled by the factor required to yield an invariant mass of \qty{125}{\GeV} to improve the mass resolution of the reconstructed charginos and neutralinos, as studied in signal MC simulation. This correction is only applied for the chargino and neutralino reconstruction. In events with two or more charged leptons, the two Higgs boson candidates and the two leading leptons are used to reconstruct two chargino candidates by pairing each Higgs boson candidate with one lepton. Both possible pairings are tested, and the pairing with the minimum mass asymmetry is selected. The mass asymmetry is defined as

\begin{equation}
\asym = \frac{|\mchinooneone - \mchinoonetwo|}{\mchinooneone + \mchinoonetwo},
\label{eq:asym}
\end{equation}

where the \mchinooneone is the mass of the chargino candidate with the higher reconstructed mass and \mchinoonetwo is the mass of the chargino candidate with the lower reconstructed mass. The alternative jet pairing with the maximum mass asymmetry is used to define the variable \mchinoonetworej, which corresponds to the lower of the two reconstructed chargino candidate masses in that ``rejected'' pairing. In events with only one charged lepton, one Higgs boson candidate is paired with the lepton to form a chargino candidate, while the other is paired with the $\vec{p}_\text{T}^\text{miss}$ (representing the neutrino in the signal model) to form a neutralino candidate. The 1-lepton channel is optimized to search for high-mass charginos and neutralinos, which tend to each decay into a back-to-back Higgs boson and lepton. The pairing that maximizes the \(\Delta R\) between the Higgs boson candidate and the charged lepton is therefore used.
As only the transverse component of the neutrino momentum is available (from $\vec{p}_\text{T}^\text{miss}$), the mass of the neutralino cannot be fully reconstructed. Instead, the neutralino's transverse mass \(\mTninoone=m_{\text{T}}(h,\vec{p}_\text{T}^\text{miss})\) is used. The transverse mass of objects \(i,\,j\) is calculated as
\begin{equation}
m_{\text{T}}(i,j) = \sqrt{m_i^2+m_j^2+2(E_{\text{T},i}E_{\text{T},j} - p_{\text{T},i}p_{\text{T},j}\cos(\Delta\phi_{ij}))},
\label{eq:mt}
\end{equation}
where \(\Delta \phi_{ij}\) is the transverse angular separation between the objects and \(E_{\text{T},i}=\sqrt{m_{i}^2+p_{\text{T},i}^2}\). The $\vec{p}_\text{T}^\text{miss}$ is assumed to come from a massless neutrino.
For both 1-lepton and 2-lepton regions, \HT is defined as the scalar sum of \pT of the four jets and 1--2 charged leptons used in the chargino and neutralino reconstruction.

\subsection{Region definitions}

Two signal regions (SRs) are defined for this analysis, SR2$\ell$ and SR1$\ell$. These regions are used to search for new physics in a simultaneous fit, described in more detail in Sections~\ref{sec:bkg} and \ref{sec:result}. Their definitions, as well as those of the control and validation regions described in Section~\ref{sec:bkg}, are shown in Table~\ref{tab:regions2l} and Table~\ref{tab:regions1l} respectively. Each region has a \RunTwo and a \RunThr counterpart, using the same definitions.

SR2$\ell$ targets $\chinoonepm\chinoonemp$ production in events with two leptons (2$\ell$). The reconstructed Higgs boson masses, \mhone and \mhtwo are required to be close to the Higgs boson mass of \qty{125}{\GeV}. The \mhtwo requirements are shifted to lower masses as muons and neutrinos from $b$-hadron decays can lead to an underestimate of the Higgs boson mass. The \HT selection is used to reject lower-energy background processes, while the requirement that the dilepton invariant mass \mll is at least \qty{15}{\GeV} higher than the $Z$ boson mass is used to reject processes with leptonic $Z$ decays including \ttZ, \Zjets, and diboson production. The \asym requirement rejects background processes by ensuring that the two reconstructed chargino masses are consistent. The pre-fit distribution of \asym in \RunTwo SR2$\ell$ without this selection is shown in Figure~\ref{fig:ac_nminus1}. The \mchinoonetworej selection reduces \ttbar backgrounds. SR2$\ell$ is binned in \mchinooneone, using eight bins with lower bin edges of $[125, 175, 225, 300, 400, 500, 600, 700]~\gev$, where the last bin includes all events with $\mchinooneone>\qty{700}{\GeV}$. The bins start at \qty{125}{\GeV} as the chargino must be heavier than the Higgs boson. The bins are narrower at low masses to capture features of the distribution and wider at high masses to avoid creating bins with very low event counts. The expected pre-fit distribution of \mchinooneone in \RunTwo SR2$\ell$ using this binning is shown in Figure~\ref{fig:mc1}.

SR1$\ell$ targets $\chinoonepm\ninoone$ production in events with one lepton (1$\ell$). This region makes use of the higher cross section relative to $\chinoonepm\chinoonemp$ to provide increased signal yields for high-mass signals. Similarly to SR2$\ell$, the two reconstructed Higgs boson candidates are required to have masses consistent with the Higgs boson mass. The \HT selection of \qty{800}{\GeV} is tighter than for SR2$\ell$, as SR1$\ell$ is optimized for higher mass signals that yield higher \HT values. The minimum transverse mass of the $\vec{p}_\text{T}^\text{miss}$ and any of the three leading \bjets is labeled \mTbmin, computed with the \bjet mass set to \qty{0}{\GeV}. For \ttbar events with a single leptonic $W$ boson decay, \mTbmin has an upper bound of \(\sqrt{m_{\Pqt}^2 - m_{\PW}^2 }\approx\qty{150}{\gev}\) and is used to further reduce the semileptonic \ttbar background. Events with low \met are rejected because signals have high \met values due to the energetic neutrino from the neutralino decay.
The transverse mass of the lepton and $\vec{p}_\text{T}^\text{miss}$ (\mTlep, see Equation~\ref{eq:mt}) is required to be above the $W$ boson mass to reject events with leptonic $W$ boson decays from processes such as \ttbar, \ttw, single-top (including \(\PW\Pqt\)), and diboson production.
SR1$\ell$ uses a single bin due to the low predicted event count. The expected pre-fit distribution of \mTninoone in \RunTwo SR1$\ell$ is shown in Figure~\ref{fig:mtn1}.

\begin{table}[t]
\centering
\caption{Summary of the signal, control, and validation regions used for the 2-lepton analysis regions targeting \chinoonepm\chinoonemp signals. All regions require at least four jets and two oppositely charged leptons. Each region has both a \RunTwo and a \RunThr version, which use the same definition.}
\label{tab:regions2l}
\resizebox{\columnwidth}{!}{
\begin{tabular}{ccccccccc}
\toprule
Region  &  \Nbjet   &    \mhone[\GeV{}]    &   \mhtwo[\GeV{}]    & \HT[\GeV{}] & \mll[\GeV{}] & \asym & \mchinoonetworej[\GeV{}] & \mchinooneone[\GeV{}] \\ \midrule
SR2$\ell$   & \(\ge 3\) &  \(\in [100, 150]\)   &  \(\in [85, 135]\)   &  \(> 400\)   &  \(> 106.2\)  &    \(<0.2\)    &         \(> 200\)         &          ---           \\ \midrule
CR2$\ell$2b  &     2     &  \(\in [100, 150]\)   &  \(\in [85, 135]\)   &     ---      &  \(> 106.2\)  &    \(<0.2\)    &         \(< 200\)         &       \(< 500\)        \\
CR2$\ell$3b  & \(\ge 3\) & \(\notin[100, 150]\)  & \(\notin [85, 135]\) &  \(> 400\)   &  \(> 106.2\)  &    \(>0.2\)    &            ---            &       \(< 700\)        \\ \midrule
VR2$\ell$2b  &     2     &  \(\in [100, 150]\)   &  \(\in [85, 135]\)   &     ---      &  \(> 106.2\)  &    \(<0.2\)    &         \(> 200\)         &       \(< 500\)        \\
VR2$\ell$3b  & \(\ge 3\) & \(\notin [100, 150]\) &  \(\in [85, 135]\)   &  \(> 400\)   &  \(> 106.2\)  &    \(>0.2\)    &            ---            &          ---           \\ \bottomrule
\end{tabular}}
\end{table}

\begin{table}[t]
\centering
\caption{Summary of the signal, control, and validation regions used for the 1-lepton analysis regions targeting \chinoonepm\ninoone signals. All regions require at least four jets and one lepton. Each region has both a \RunTwo and a \RunThr version, which use the same definition.}
\label{tab:regions1l}
\resizebox{\columnwidth}{!}{%
\begin{tabular}{cccccccc}
\toprule
Region  &  \Nbjet   &    \mhone[\GeV{}]     &   \mhtwo[\GeV{}]    & \HT[\GeV{}] & \mTbmin[\GeV{}] & \MET[\GeV{}] & \mTlep[\GeV{}] \\ \midrule
SR1$\ell$   & \(\ge 3\) &   \(\in [100, 150]\)   &  \(\in [85, 135]\)   &  \(> 800\)   &     \(> 80\)     &   \(> 150\)   &    \(> 100\)    \\ \midrule
CR1$\ell$3b  & \(\ge 3\) & \(\notin[100, 150] \)  & \(\notin [85, 135]\) &  \(> 800\)   &     \(> 80\)     &   \(> 150\)   &  \(\leq 100\)   \\ \midrule
VR1$\ell$3b1 & \(\ge 3\) & \(\notin [100, 150]\)  &  \(\in [85, 135]\)   &  \(> 800\)   &     \(> 80\)     &   \(> 150\)   &  \(\leq 100\)   \\
VR1$\ell$3b2 & \(\ge 3\) & \(\notin [100, 150] \) & \(\notin [85, 135]\) &  \(> 800\)   &     \(> 80\)     &   \(> 150\)   &    \(> 100\)    \\ \bottomrule
\end{tabular}}
\end{table}

\begin{figure}[h!]
\centering
\subfloat[]{
\centering
\includegraphics[width=0.49\linewidth]{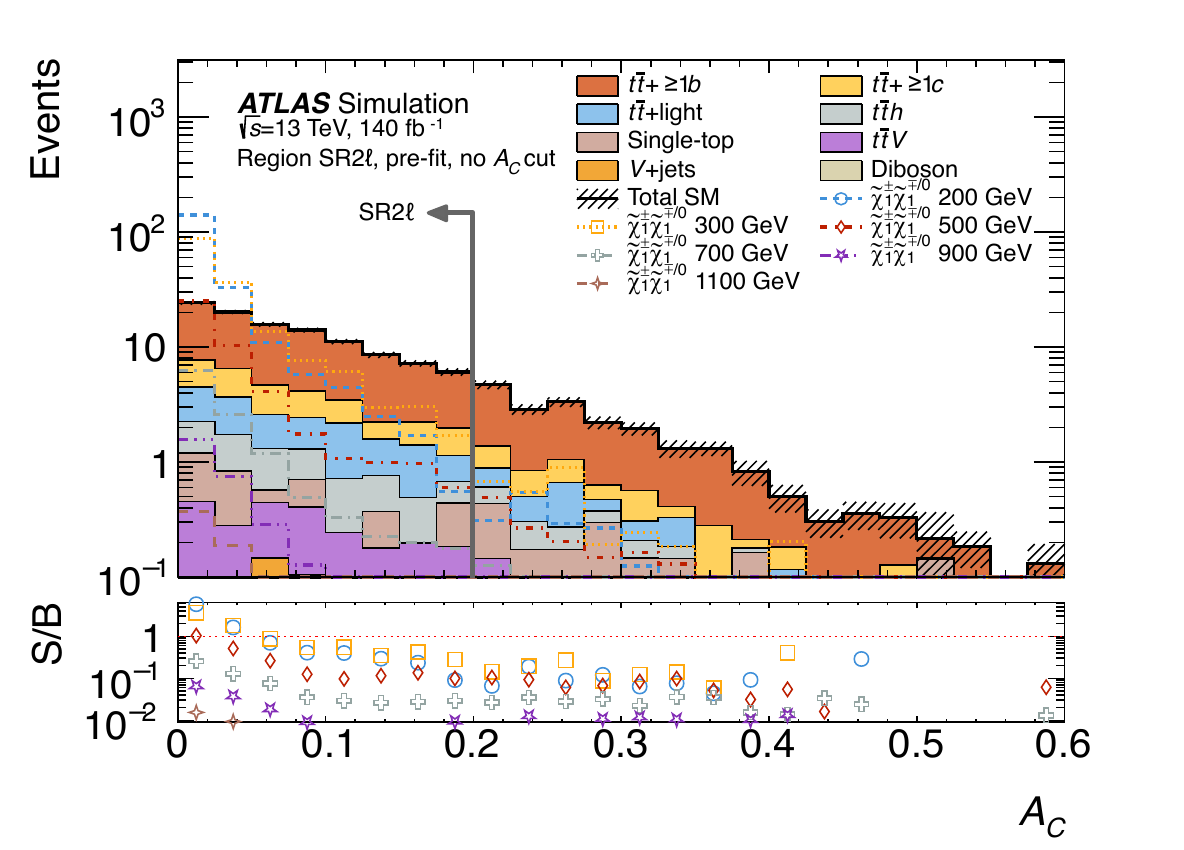}
\label{fig:ac_nminus1}
}~
\subfloat[]{
\centering
\includegraphics[width=0.49\linewidth]{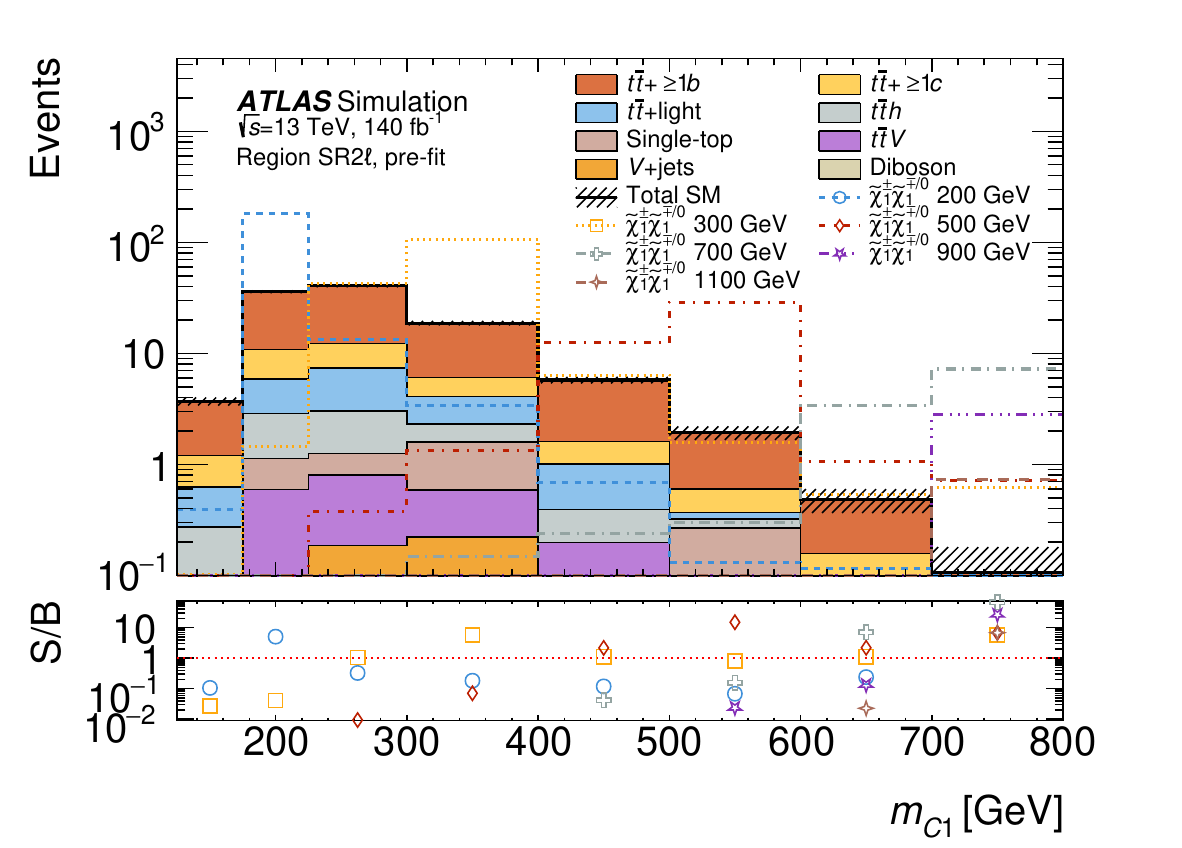}
\label{fig:mc1}
}\\
\subfloat[]{
\centering
\includegraphics[width=0.49\linewidth]{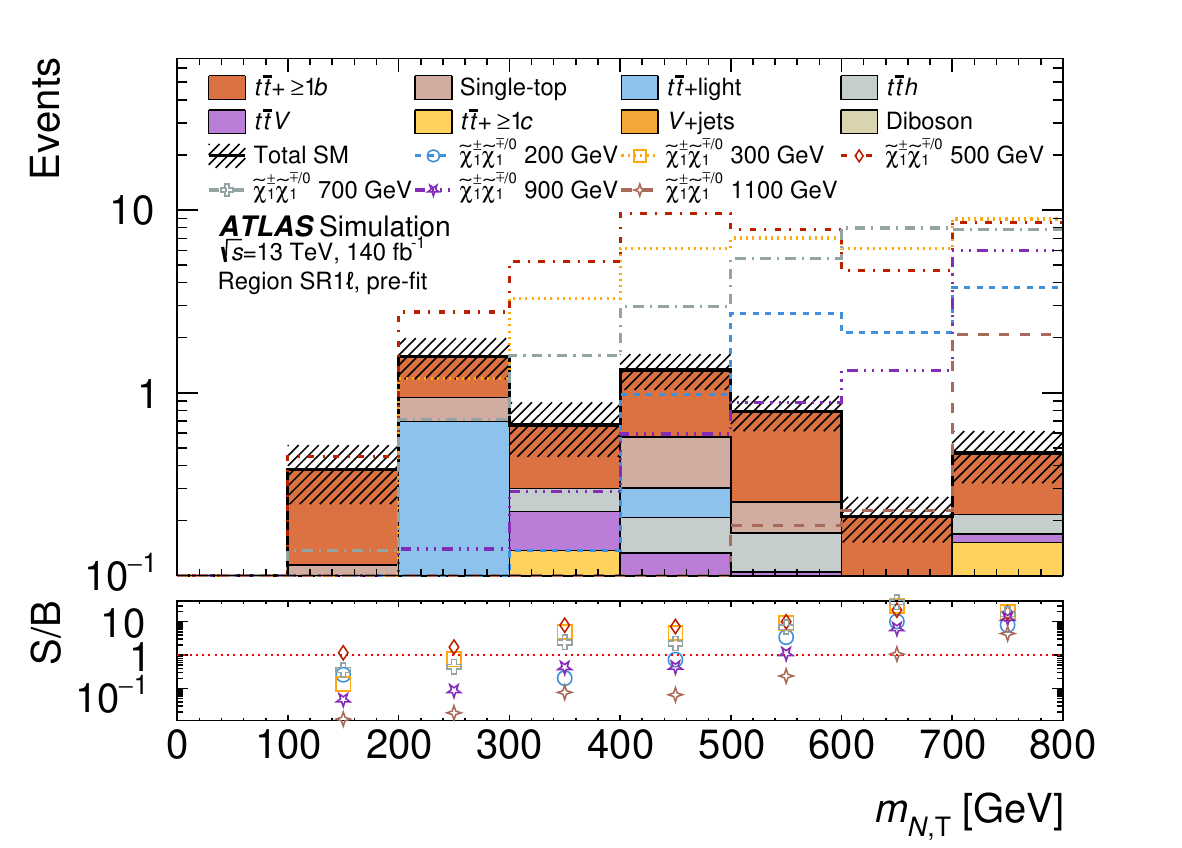}
\label{fig:mtn1}
}
\caption{Expected \RunTwo pre-fit distributions for signal and background samples for (a) the mass asymmetry \asym in SR2$\ell$ before the \asym selection, (b) the higher mass of the two reconstructed chargino candidates \mchinooneone in SR2$\ell$, and (c) the transverse mass of the neutralino \mTninoone in SR1$\ell$. The lower panels show the signal to background ratios. The hatching shows the MC statistical uncertainty in the predicted background yields. The sum of the signal contributions from $\chinoonepm\chinoonemp$ and $\chinoonepm\ninoone$ is shown, assuming $\mathcal{B}(\chinoonepm\rightarrow h\ell^\pm)=100\%$ and $\mathcal{B}(\ninoone\rightarrow h\nu)=100\%$, with equal decay branching fractions to each lepton flavor. The generator-level signal masses are shown in the legend. Overflow is included in the last bin.}
\label{fig:mc1_mtn1}
\end{figure}

Four regions, referred to as ``discovery regions'',  are used to obtain results on new physics, independent of a specific signal hypothesis. The first two, SR2$\ell$ and SR1$\ell$, are single-bin regions matching the definitions for the SRs in Tables~\ref{tab:regions2l} and \ref{tab:regions1l}. In addition, SR2$\ell$700 is defined as SR2$\ell$ with an additional requirement of $\mchinooneone>\qty{700}{\GeV}$ and is equivalent to the last bin of SR2$\ell$. Finally, SR1$\ell$600 is defined as SR1$\ell$ with an additional requirement on the transverse mass of the neutralino $\mTninoone>\qty{600}{\GeV}$. Each of these four regions is defined for both \RunTwo and \RunThr as they cannot be combined in a model-independent way due to the different center-of-mass energies.

% End of text imported from the .//sections/event_selection.tex input file

%
\section{Background estimation}
\label{sec:bkg}

% The next lines are included from the .//sections/background_estimation.tex input file
The background is estimated by using the MC simulation samples described in Section~\ref{sec:data_mc_samples}. The main backgrounds are from \ttbar (\ttb, \ttc, and \ttlight) production, and are normalized to data in control regions (CRs). Separate normalizations are used for the 1-lepton and 2-lepton regions due to the different phase space. The largest contribution in the SRs is given by \ttb production. Smaller backgrounds that are present in this search are \tth, \ttz, \ttw, single-top, \Zjets, \Wjets, and diboson production. These backgrounds, referred to as minor backgrounds, collectively constitute less than 10\% of the pre-fit expected background in SR2$\ell$ and less than a third of the pre-fit expected background in SR1$\ell$.

Two CRs, CR2$\ell$2b and CR2$\ell$3b, are used to measure the normalizations of dileptonic \ttbar backgrounds. Separate normalizations are determined for \RunTwo and \RunThr due to differences in the experimental setup and simulations. These CRs are defined in Table~\ref{tab:regions2l}. CR2$\ell$3b is constructed by inverting the SR2$\ell$ selections on \mhone, \mhtwo, and \asym to select a region with low signal contribution while removing the requirements on \mchinoonetworej to further reduce signal contamination and lower the statistical uncertainty. An additional requirement is placed on \mchinooneone to prevent contributions from signals with large chargino masses.
This region is used to measure a common normalization factor for \ttb and \ttc production, referred to as \ttHF (heavy-flavor). MC simulation predicts similar relative contributions from \ttb and \ttc in CR2$\ell$3b and SR2$\ell$, allowing a joint measurement with smaller statistical uncertainty. The \tthf purity in this region is over 80\%. The expected signal contributions are found to be less than 8\% of the pre-fit expected background for any signal mass hypothesis.

CR2$\ell$2b is used to measure the \ttlight normalization. This uses events with only two \bjets, as \ttbb events have four true \bjets and \ttc events are more likely than \ttlight events to have a third reconstructed \bjet due to the higher charm mis-tagging rate compared to light jets. Contributions from signals are limited by inverting the SR2$\ell$ selection on \mchinoonetworej and removing the requirement on \HT. An upper bound is placed on \mchinooneone to prevent contributions from signals with large chargino masses. CR2$\ell$2b has a \ttlight purity of approximately 80\%. The expected signal contributions are found to be less than 2\% of the pre-fit expected background for any signal mass hypothesis.

CR1$\ell$3b is used to measure the normalization of semileptonic \ttbar backgrounds. Separate normalizations are determined for \RunTwo and \RunThr. This measures a common normalization factor for all semileptonic \ttbar backgrounds (\ttb, \ttc, and \ttlight), jointly referred to as \ttonel. These are measured jointly to lower the statistical uncertainty, as the \ttc and \ttlight backgrounds are expected to contribute a total of less than one event in SR1$\ell$ and therefore do not require separate measurements. The dominant background component in SR1$\ell$ and CR1$\ell$3b is \ttb. CR1$\ell$3b is defined in Table~\ref{tab:regions1l} and is constructed by inverting the SR1$\ell$ requirements on \mhone, \mhtwo, and \mTlep to reduce signal contributions. The \ttonel purity in this region is over 80\%, while the expected signal contributions are found to be less than 1\% of the pre-fit expected background for any signal mass hypothesis.

Validation regions (VRs) are constructed to test the extrapolation of the background model from the CRs to the SRs. These regions are defined to fall in the phase space between their corresponding CR and SR, as shown in Tables~\ref{tab:regions2l} and \ref{tab:regions1l}. VR2$\ell$2b and VR2$\ell$3b test the extrapolations over \mchinoonetworej and \mhtwo respectively, while VR1$\ell$3b1 and VR1$\ell$3b2 test the extrapolations over \mhtwo and \mTlep respectively. The expected signal contributions are found to be less than 12\% of the pre-fit expected background in each VR for any signal mass hypothesis.

The agreement between data and predicted backgrounds for different variables are checked in each CR and VR except VR1$\ell$3b2. This region is only checked in a single bin as it has high signal contamination in some kinematic regimes. Reconstructed chargino and Higgs boson masses are shown for the VRs in Figure~\ref{fig:datamc_kinematics}, where good modeling of the SM background is seen.

\begin{figure}[h!]
\centering
\subfloat[]{
\centering
\includegraphics[width=0.49\linewidth]{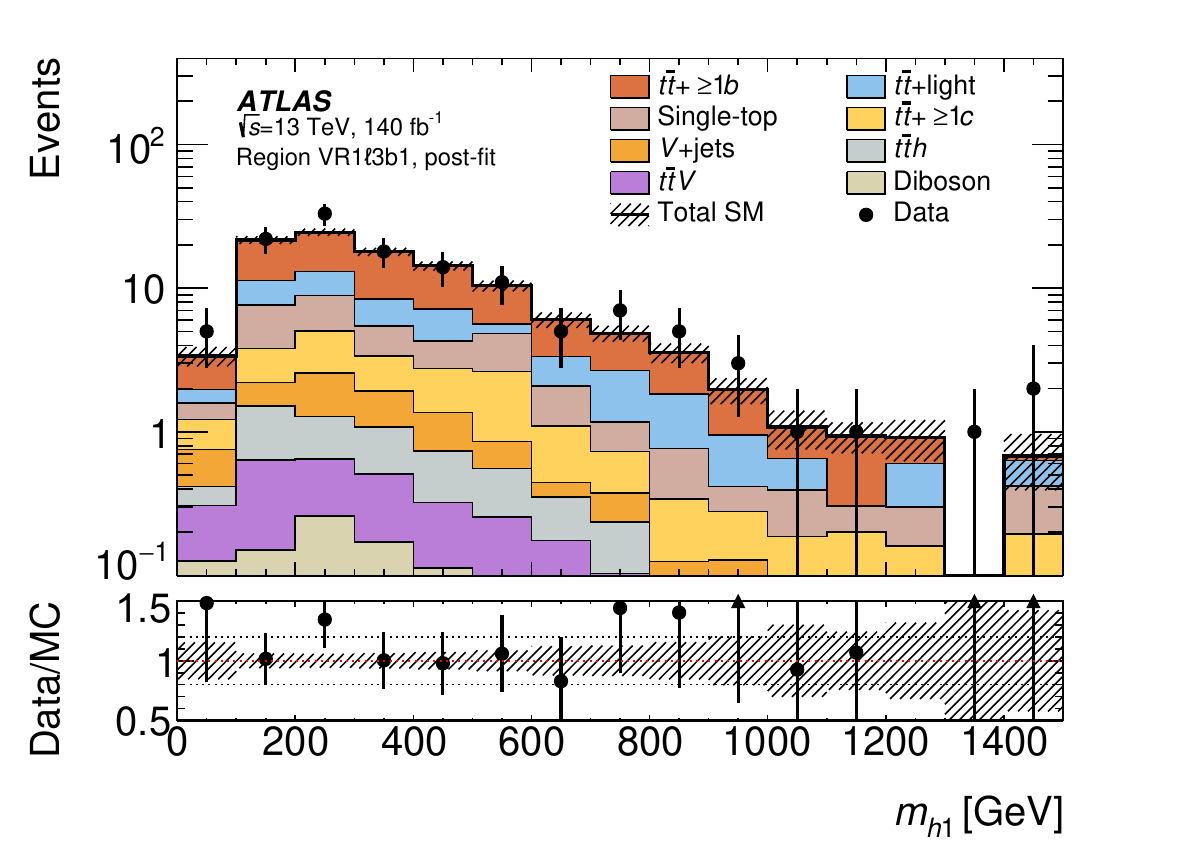}
\label{fig:datamc_vr1l3b1_mh1}
}~
\subfloat[]{
\centering
\includegraphics[width=0.49\linewidth]{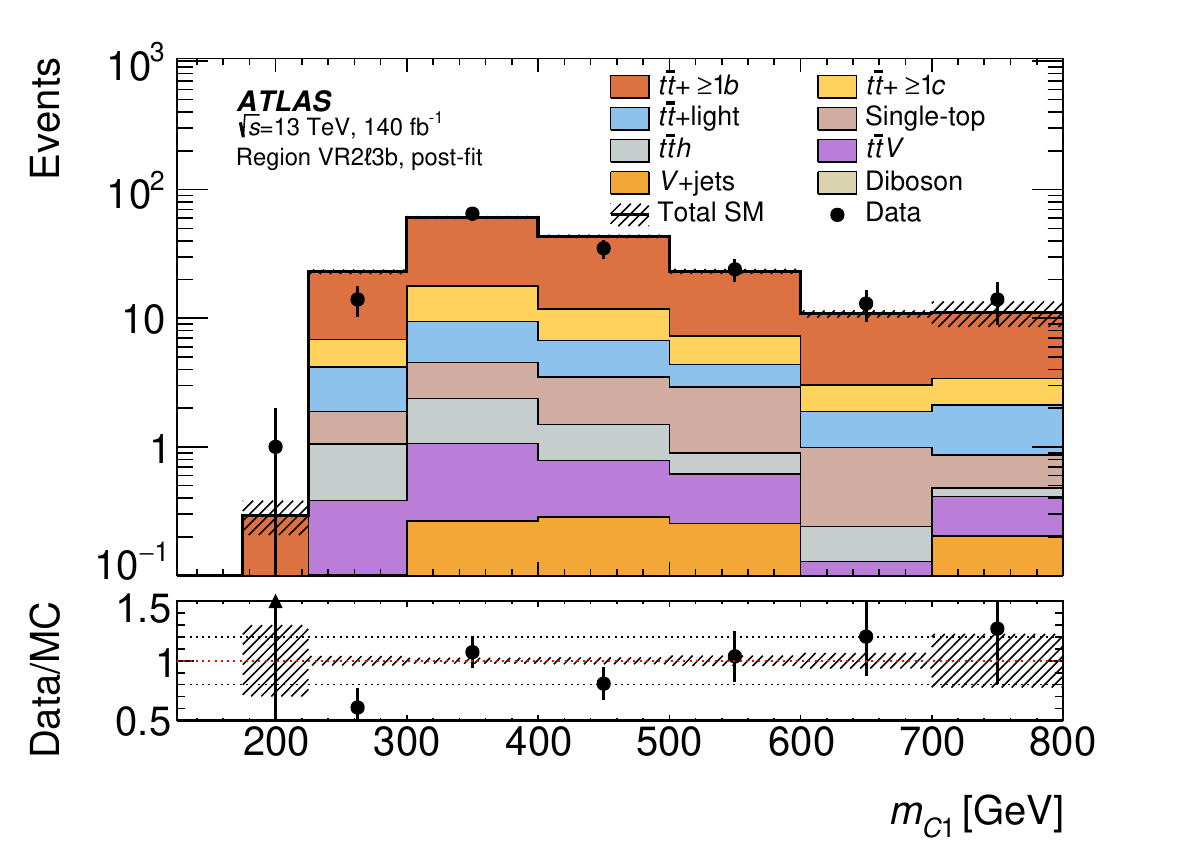}
\label{fig:datamc_vr2l3b_mc1}
}\\
\subfloat[]{
\centering
\includegraphics[width=0.49\linewidth]{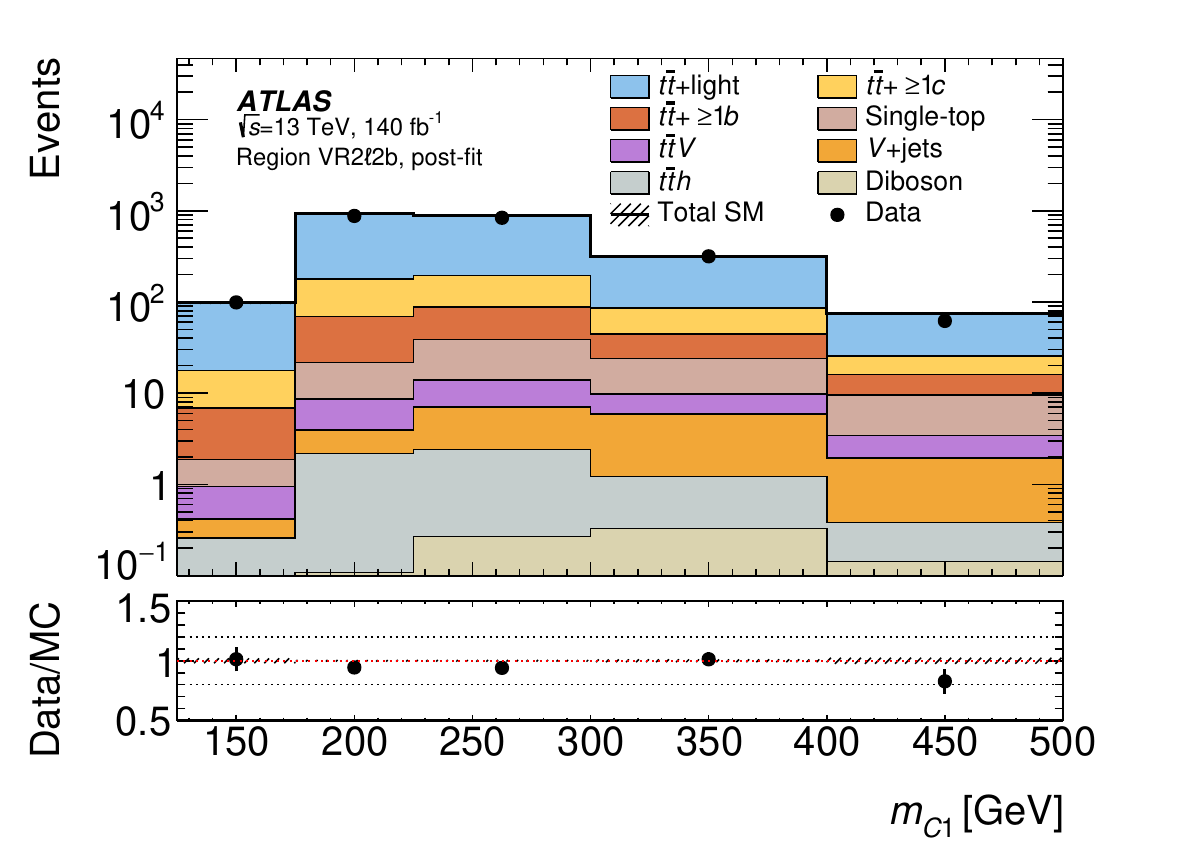}
\label{fig:datamc_vr2l2b_mc1}
}
\subfloat[]{
\centering
\includegraphics[width=0.49\linewidth]{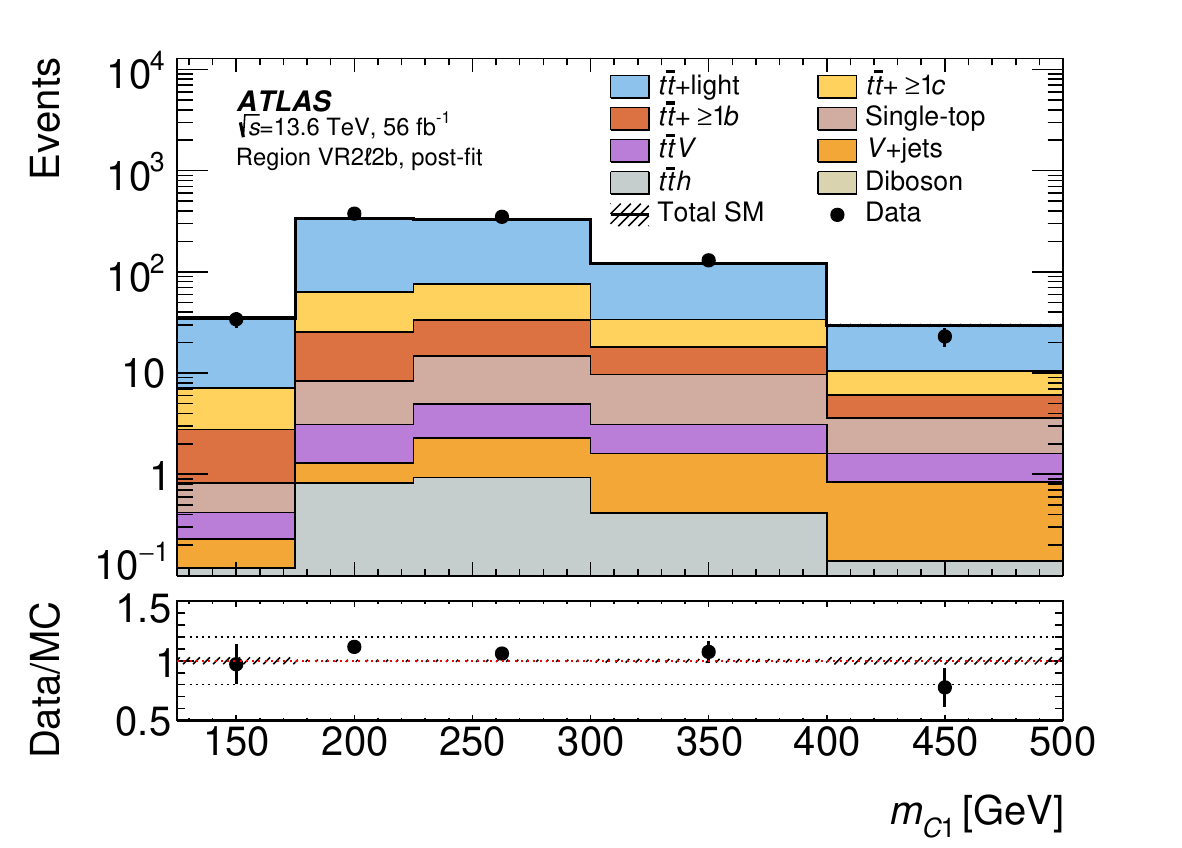}
\label{fig:datamc_vr2l2b_mc1_run3}
}
\caption{Data and background distributions in the validation regions for (a) the \mhone distribution in \RunTwo VR1$\ell$3b1, (b) the \mchinooneone distribution in \RunTwo VR2$\ell$3b, (c) the \mchinooneone distribution in \RunTwo VR2$\ell$2b, and (d) the \mchinooneone distribution in \RunThr VR2$\ell$2b. The backgrounds are calculated using pre-fit values but with the normalization factors measured from the CR background-only fit applied. The hatching shows the MC statistical uncertainty in the predicted background yields. The lower panels show the ratio of data to predicted background. Overflow is included in the last bin.}
\label{fig:datamc_kinematics}
\end{figure}

\subsection{Background-only fit results}

Profile-likelihood fits are performed to measure the agreement between data and predicted background.
There are six free-floating background normalization factors in total. Each Run has two normalization factors for \tthf and \ttlight flavor production in 2-lepton regions and one normalization factor for \ttbar production in 1-lepton regions. The MC statistical uncertainties and the systematic uncertainties described in Section~\ref{sec:systematics} are treated as nuisance parameters.
Unless otherwise noted, all fits use these six normalization factors and both 1-lepton and 2-lepton regions in both \RunTwo and \RunThr. All fits are performed using the \textsc{pyhf} framework~\cite{pyhf, pyhf_joss}.

A background-only fit is performed in the CRs using a single bin for each region. The measured normalization factors are shown in Table~\ref{tab:normfactors}.

\begin{table}[htb!]
\centering
\caption{Fit results for normalization factors from the background-only fit in the CRs. The first listed uncertainties are from the fit using all systematic and statistical uncertainties, while the values in parentheses are determined from the fit using only statistical uncertainties. Due to correlations between the normalization factors and systematic uncertainties, the relative uncertainties in the predicted backgrounds yields may be smaller than those on the normalization factors.}
\label{tab:normfactors}
\begin{tabular}{lcc}\toprule
Process (Channel) & \RunTwo Norm. Factor & \RunThr Norm. Factor \\ \midrule
\tthf (2$\ell$) &1.16$\pm$0.46 ($\pm$0.06 stat.) & 1.08$\pm$0.42 ($\pm$0.10 stat.)\\
\ttlight (2$\ell$) &0.93$\pm$0.20 ($\pm$0.03 stat.) & 0.85$\pm$0.20 ($\pm$0.05 stat.)\\
\ttbar (1$\ell$) &0.88$\pm$0.33 ($\pm$0.06 stat.) & 1.01$\pm$0.45 ($\pm$0.08 stat.)\\
\bottomrule
\end{tabular}
\end{table}

Results from the background-only fit are extrapolated to the VRs, with the agreement between data and the post-fit background prediction shown in Figure~\ref{fig:datamc_vr}. No significant mismodeling is observed. The largest deviation in yield between data and predicted background is $1.2\sigma$, which is seen in \RunThr VR2$\ell$2b.

\begin{figure}[h!]
\centering
\includegraphics[width=0.49\linewidth]{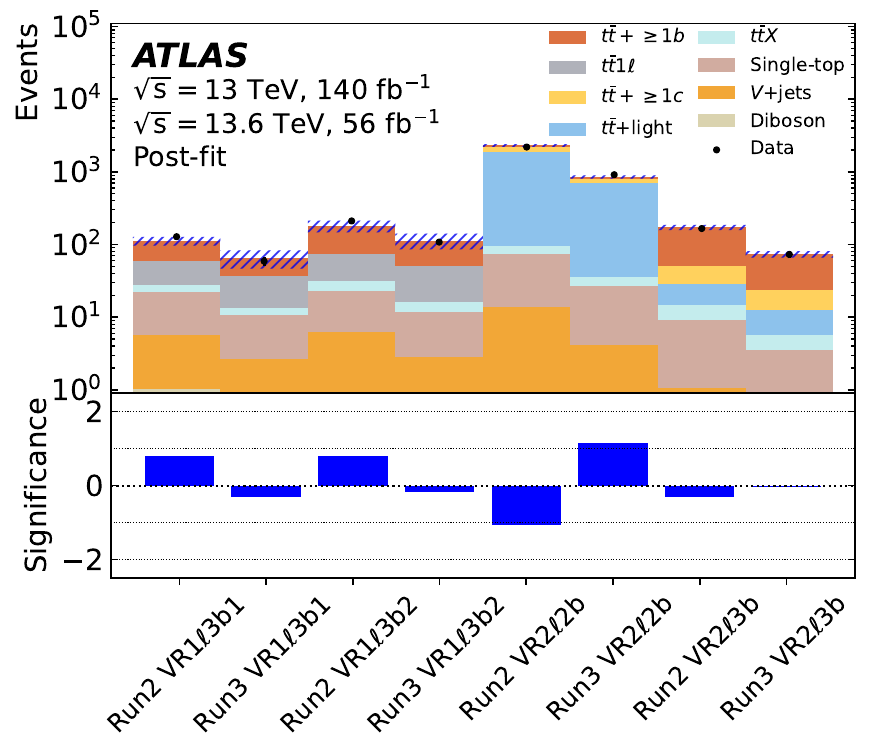}
\caption{Data and post-fit background yields as determined from the background-only fit extrapolated to the validation regions. The hatching shows the uncertainty in the predicted yields from MC statistical and systematic effects. The lower panel shows the significance of the deviations between data and predicted background. The \ttonel category includes all \ttc and \ttlight production with one reconstructed lepton. \(\ttbar X\) includes \tth, \ttZ, and \ttW production. Significances are calculated following the method in Ref.~\cite{ATL-PHYS-PUB-2020-025}.}
\label{fig:datamc_vr}
\end{figure}

% End of text imported from the .//sections/background_estimation.tex input file

%
\section{Systematic uncertainties}
\label{sec:systematics}

% The next lines are included from the .//sections/systematics.tex input file
Predicted signal and background MC simulation yields are subject to uncertainties. Three types of uncertainties are applied: experimental uncertainties, theoretical uncertainties, and MC statistical uncertainties.

Experimental uncertainties account for systematic effects in ATLAS calibration procedures and efficiency measurements and are applied to all signal and background MC simulations. Experimental jet uncertainties are applied on the jet energy scale and resolution~\cite{PERF-2014-07, PERF-2015-09, JETM-2018-05}, JVT~\cite{PERF-2014-03} and fJVT~\cite{PERF-2016-06}, and flavor tagging efficiencies~\cite{FTAG-2023-05}. Experimental muon uncertainties include uncertainties on the muon isolation, reconstruction, %
and trigger efficiency correction factors~\cite{MUON-2022-01, MUON-2018-03, PERF-2015-10, TRIG-2018-01, TRIG-2022-01}. Additional systematic uncertainties affect the muon momenta due to variations in the  momentum scale, track resolution, and charge-dependent corrections. For electrons, experimental uncertainties are applied on the identification, isolation, reconstruction, and trigger efficiency correction factors, as well as on the energy scale and resolution~\cite{PERF-2013-05, EGAM-2021-01, TRIG-2018-05, TRIG-2022-01}. Scale and resolution uncertainties are applied on the $\met$ soft term~\cite{JETM-2020-03, Balunas:2023fpu}. Further experimental uncertainties considered are on the pileup conditions~\cite{STDM-2015-05} and luminosity~\cite{DAPR-2021-01, ATL-DAPR-PUB-2024-001}. Luminosity uncertainties are not applied to \ttbar and \ttbb backgrounds as they rely on data-driven normalization measurements. Uncertainties related to electrons, $\met$, and most of the jet energy scale and resolution components are correlated between \RunTwo and \RunThr, while the remaining uncertainties are not correlated between Runs. The largest experimental uncertainties are related to the jet energy resolution. As measured using the background-only fit extrapolated to the signal regions, these are collectively 1.3\% (3.5\%) of the post-fit yields in \RunTwo (\RunThr) SR2$\ell$ and 23\% (35\%) of the post-fit yields in the \RunTwo (\RunThr) SR1$\ell$.

Theoretical uncertainties are applied to the background and signal MC samples. These uncertainties are evaluated by comparing simulations using nominal values of theory parameters with alternative or reweighted samples using variations on these parameters. Uncertainties in the matrix element, parton shower model, factorization and renormalization scales, modeling of initial and final-state radiation, parton distribution functions, and strong coupling constant are applied to the \ttb, \ttc, and \ttlight backgrounds. The matrix element and parton shower uncertainties are calculated by comparing the nominal samples to samples generated with the \PYTHIA[8] \pthard parameter set to 1.0 and with the \HERWIG[7] parton shower model~\cite{Bahr:2008pv,Bellm:2015jjp,Bellm:2017jjp,Bellm:2019zci} respectively. In addition, uncertainties on the \hdamp parameter are applied to the \ttc and \ttlight backgrounds by comparing the nominal samples to samples produced with the \hdamp parameter set to 3.0~\mtop. A 50\% normalization uncertainty is applied to each of the remaining background samples (\tth, \ttV, single-\Pqt, \Vjets, and diboson). As the background is dominated by \ttbar and \ttbb and the analysis is statistically limited, the uncertainties on these minor backgrounds have no significant impact on the results. Background theoretical uncertainties are correlated between \RunTwo and \RunThr, but are not correlated between 1-lepton and 2-lepton regions due to the different phase space. The largest theoretical uncertainties in SR2$\ell$ are from the parton shower model. As measured using the background-only fit extrapolated to the signal regions, these are collectively 4.2\%--6.0\% of the post-fit yields in the 2-lepton SRs, depending on the Run. The largest theoretical uncertainties in SR1$\ell$ are from the modeling of final-state radiation and the parton shower modeling. As measured using the background-only fit extrapolated to the signal regions, these are each 3\%--20\% of the post-fit yields in the 1-lepton SRs, depending on the uncertainty and Run. For signal samples, uncertainties on the factorization and renormalization scales, merging scale, parton shower tuning, parton distribution function, and \alphas are considered. The factorization and renormalization scale uncertainties are the largest of these, ranging from 6.1\%--11\% for $\chinoonepm\chinoonemp$ signals in SR2$\ell$ and 8.4\%--12.6\% for $\chinoonepm\ninoone$ signals in SR1$\ell$, depending on the mass point.

The effects of different sources of uncertainties on the SR2$\ell$ and SR1$\ell$ post-fit background yields are shown in Figure~\ref{fig:bkgsyst}. Background theory uncertainties are larger than experimental uncertainties in \RunTwo, while for \RunThr experimental uncertainties are larger than theoretical uncertainties in SR1$\ell$ and some bins of SR2$\ell$. MC statistical uncertainties are smaller than both background theory and experimental uncertainties. SR1$\ell$ and the high-\mchinooneone bins in SR2$\ell$ have low MC event statistics, making the systematic uncertainty estimates prone to statistical fluctuations.

\begin{figure}[h!]
\centering
\subfloat[]{
\centering
\includegraphics[width=0.49\linewidth]{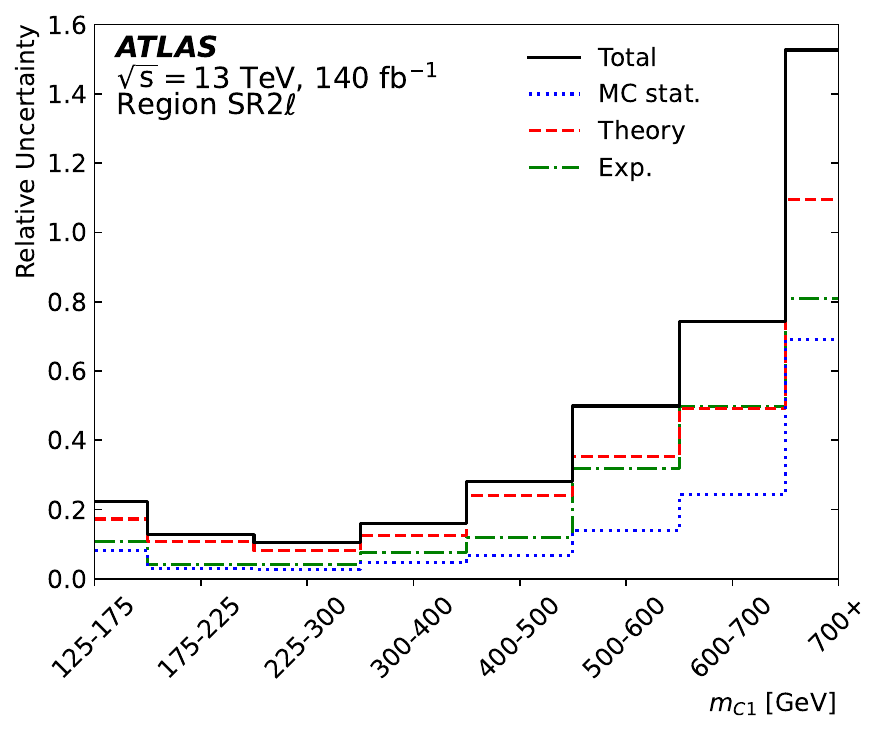}
\label{fig:bkgsyst_sr2l_run2}
}~
\subfloat[]{
\centering
\includegraphics[width=0.49\linewidth]{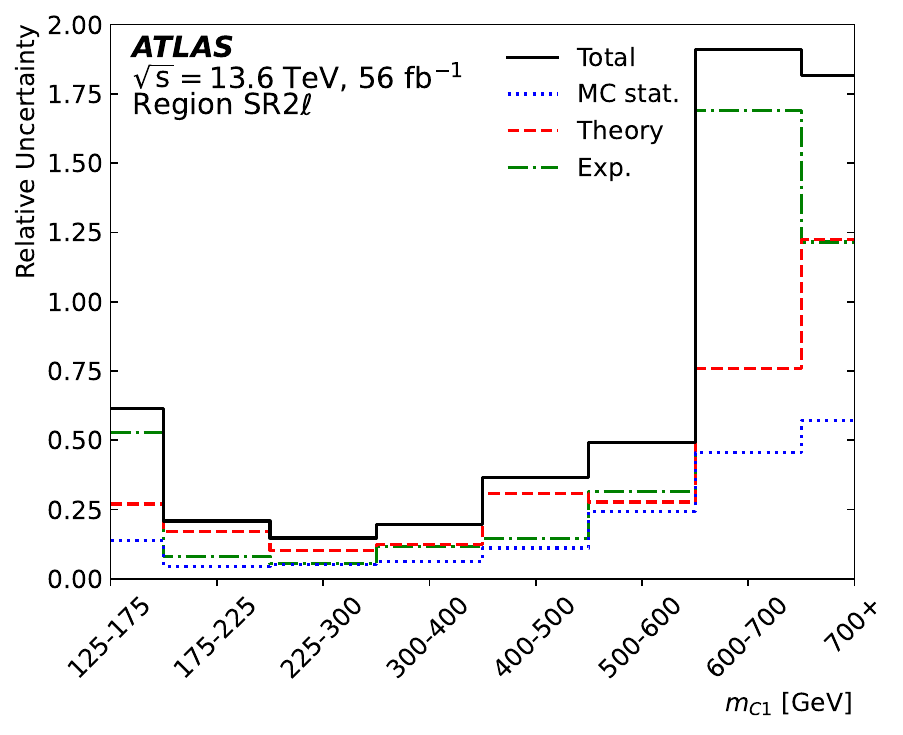}
\label{fig:bkgsyst_sr2l_run3}
}\\
\subfloat[]{
\centering
\includegraphics[width=0.49\linewidth]{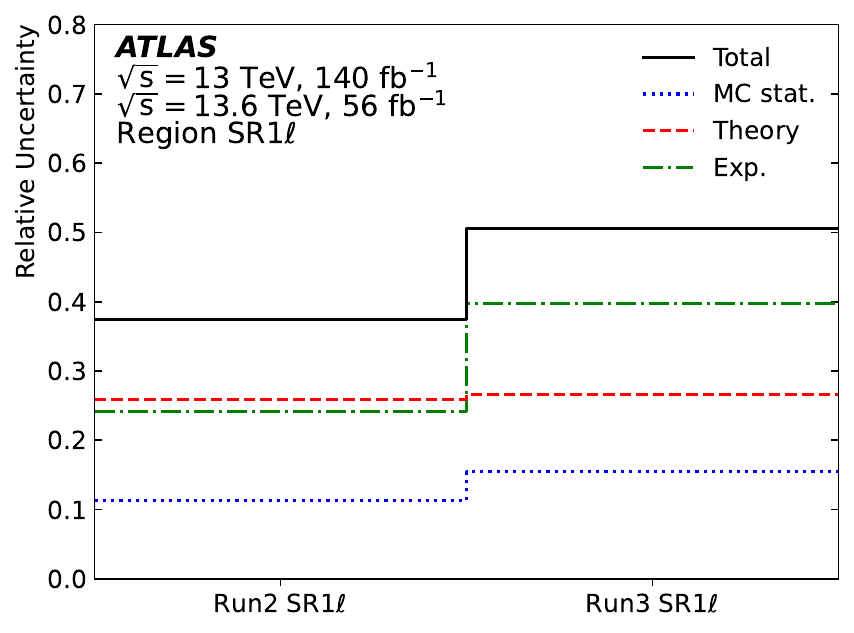}
\label{fig:bkgsyst_sr1l}
}
\caption{Background uncertainties as a fraction of total post-fit yields as determined from the background-only fit extrapolated to the signal regions for (a) \RunTwo SR2$\ell$, (b) \RunThr SR2$\ell$, and (c) \RunTwo and \RunThr SR1$\ell$. Exp.~stands for the experimental uncertainty while MC stat.~stands for MC statistical uncertainty. The total uncertainty includes systematic and MC statistical uncertainties. The total uncertainty is not necessarily equal to the quadrature sum of the individual uncertainties due to correlations.}
\label{fig:bkgsyst}
\end{figure}

% End of text imported from the .//sections/systematics.tex input file

%
\section{Results}
\label{sec:result}

% The next lines are included from the .//sections/results.tex input file
Results are obtained using three types of fits. First, the results of the background-only fit described in Section~\ref{sec:bkg} are extrapolated to the SRs to show the level of agreement between data and predicted background. Second, fits are performed in each discovery region to search for evidence for and set constraints on new physics in a model-independent manner. Third, an exclusion fit is performed in the SRs to set limits on the simplified model described in Section~\ref{sec:intro}.

\subsection{Background-only fit results}

Results from the background-only fit performed in the CRs (described in Section~\ref{sec:bkg}) are extrapolated to the SRs. The agreement between data and the post-fit background prediction is shown for SR1$\ell$ and each bin of SR2$\ell$ in Figure~\ref{fig:datamc_sr}. Note that the discovery region SR2$\ell$700 is defined as the last bin of SR2$\ell$. No significant excess is observed. The largest excess observed is in the $\qty{125}{\gev}$--$\qty{175}{\gev}$ bin of \RunThr SR2$\ell$, with four observed events and $1.24\pm0.76$ predicted background events. This corresponds to a local significance of $1.5\sigma$. The largest deficit occurs in the $\qty{400}{\gev}$--$\qty{500}{\gev}$ bin of \RunThr SR2$\ell$, with zero observed events and $2.58\pm0.94$ predicted background events. This corresponds to a local significance of $-2.1\sigma$.

\begin{figure}[h!]
\centering
\subfloat[]{
\centering
\includegraphics[width=0.49\linewidth]{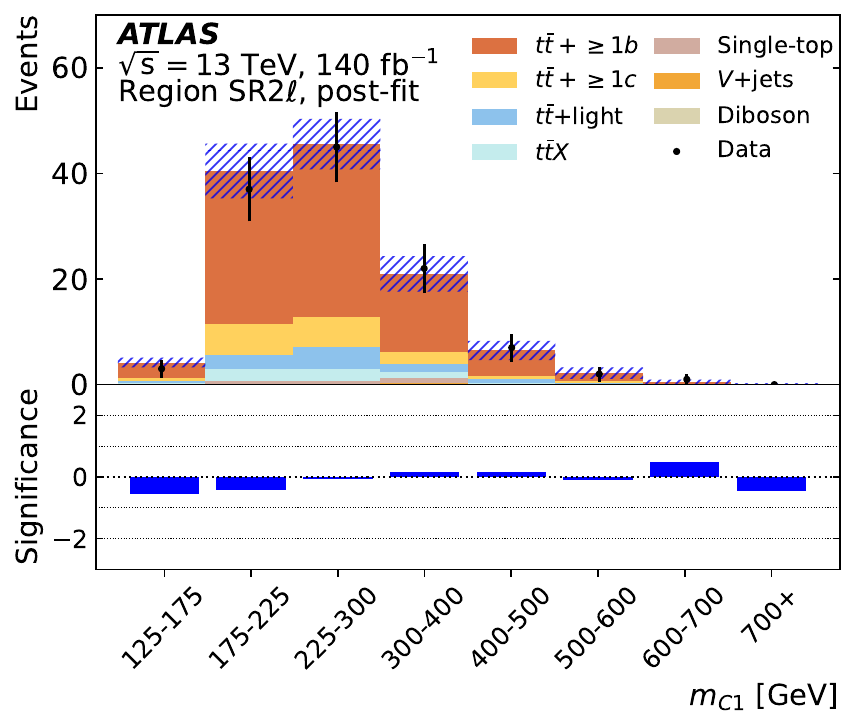}
\label{fig:datamc_sr2l_run2}
}~
\subfloat[]{
\centering
\includegraphics[width=0.49\linewidth]{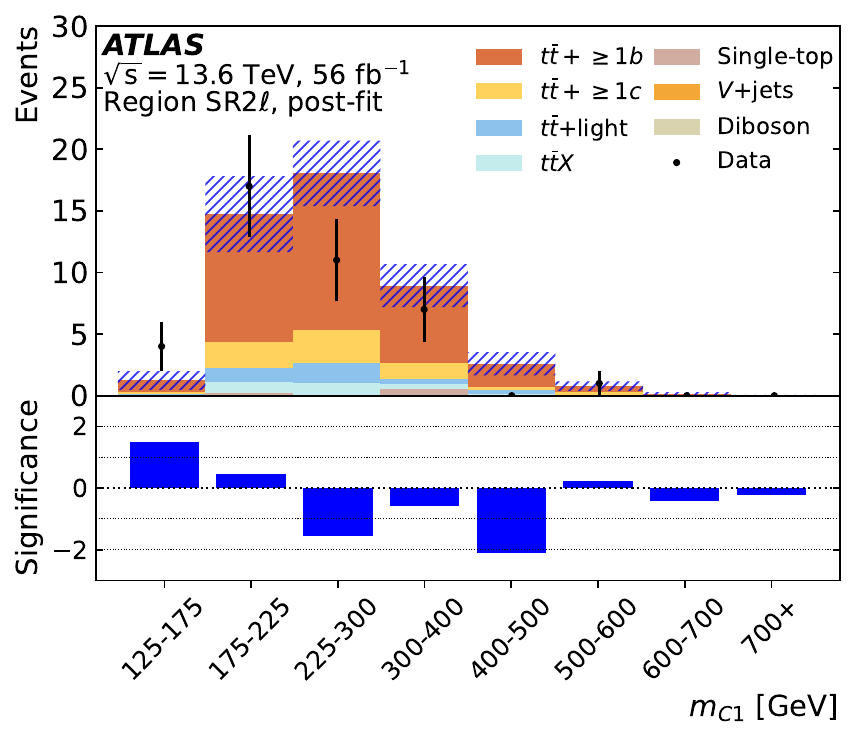}
\label{fig:datamc_sr2l_run3}
}\\
\subfloat[]{
\centering
\includegraphics[width=0.49\linewidth]{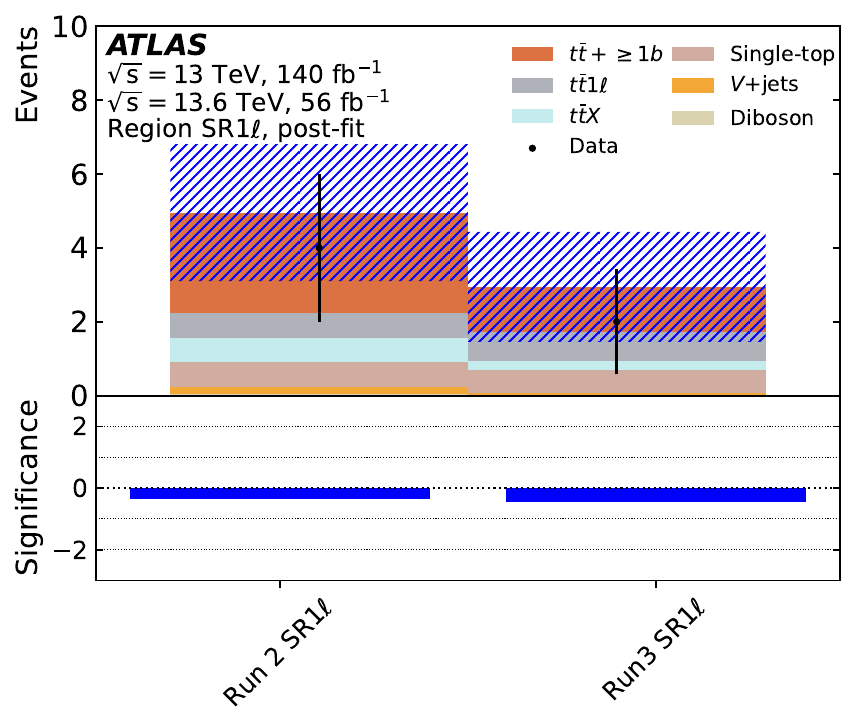}
\label{fig:datamc_sr1l}
}
\caption{Data and post-fit background yields as determined from the background-only fit extrapolated to the signal regions for (a) \RunTwo SR2$\ell$, (b) \RunThr SR2$\ell$, and (c) \RunTwo and \RunThr SR1$\ell$. The hatching shows the uncertainty in the predicted yields from MC statistical and systematic effects. The lower panel shows the significance of the deviations between data and predicted background. The \ttonel category includes all \ttc and \ttlight production with one reconstructed lepton. \(\ttbar X\) includes \tth, \ttZ, and \ttW production. Significances are calculated following the method in Ref.~\cite{ATL-PHYS-PUB-2020-025}.}
\label{fig:datamc_sr}
\end{figure}

\subsection{Model-independent results}

Model-independent results are obtained for each discovery region of the analysis. A separate fit is performed for each discovery region. For a given fit, CRs and normalization factors corresponding to a different Run or number of leptons than the discovery region are not used. Significances and upper limits are determined respectively using the $q_0$ and $\tilde{q}_\mu$ test statistics~\cite{Cowan:2010js} based on the profile likelihood ratio. The upper limits are set at 95\% confidence level (CL) using the \CLs prescription~\cite{Read:2002hq}. Regions with non-zero observed events use 50,000 pseudoexperiments to calculate $p$-values, limits, and the confidence level observed for the background-only hypothesis, CL$_\text{b}$. Regions with zero observed events use 100,000 pseudoexperiments. It is assumed that no signal events enter the CRs. %
Results are shown in Table~\ref{tab:disc_results}. No significant excess is observed, and limits are set on the yield and cross section of beyond-the-SM physics processes.

\begin{table}[t]
\caption{Results for the model-independent discovery regions. The second column shows the number of observed events ($N_\text{obs}$), while the third column shows the post-fit number of predicted events obtained using a background-only fit ($N_\text{pred}$). The fourth column shows the 95\% CL observed upper limit on the efficiency times cross section of beyond-the-SM processes ($\langle\epsilon{\sigma}\rangle_{\textrm{obs}}^{95}$[fb]). The fifth and sixth columns show the 95\% CL observed ($S^{95}_\text{obs}$) and expected ($S^{95}_\text{exp}$) upper limits on the number of events from processes beyond the SM, with the $\pm1\sigma$ uncertainty included with the expected limit. Uncertainties in $S^{95}_\text{exp}$ less than 0.05 are reported as 0.0 due to rounding. The seventh columns shows CL$_\text{b}$, the confidence level observed for the background-only hypothesis. The final column shows the discovery $p$-value ($p(s=0)$) and its significance. The $p$-value is not calculated but instead set to 0.5 for regions with a deficit relative to the post-fit background prediction. SR2$\ell$ and SR1$\ell$ are the same regions used for exclusions.}
\label{tab:disc_results}
\begin{center}
\begin{tabular}{lccccccc}
\toprule
Region  & $N_\text{obs}$ & $N_\text{pred}$ & $\langle\epsilon{\sigma}\rangle_{\textrm{obs}}^{95}$[fb] & $S_{\textrm{obs}}^{95}$ & $S_{\textrm{exp}}^{95}$ & CL$_\text{b}$ & $p(s=0)$ \\ \midrule
\RunTwo SR2$\ell$ & 117 & 120$\pm$10 & 0.24 & 33 & $ { 31 }^{ +12 }_{ -8 }$ & $0.58$ & $0.50$~$(0.00)$ \\
\RunTwo SR2$\ell$700 & 0 & 0.11$\pm$0.17 & 0.022 & 3.1 & $ { 3.0 }^{ +0.1 }_{ -0.0 }$ & $0.08$ & $0.50$~$(0.00)$ \\
\RunTwo SR1$\ell$    & 4 & 5.0$\pm$1.9 & 0.048 & 6.7 & $ { 6.7 }^{ +2.6 }_{ -1.6 }$ & $0.50$ & $0.50$~$(0.00)$ \\
\RunTwo SR1$\ell$600 & 1 & 0.62$\pm$0.51 & 0.031 & 4.3 & $ { 3.3 }^{ +1.3 }_{ -0.1 }$ & $0.78$ & $0.13$~$(1.14)$ \\ \midrule
\hline
\RunThr SR2$\ell$    & 40 & 46.4$\pm$6.1 & 0.27 & 15 & $ { 18.4 }^{ +7.6 }_{ -5.2 }$ & $0.30$ & $0.50$~$(0.00)$ \\
\RunThr SR2$\ell$700 & 0 & 0.023$\pm$0.041 & 0.053 & 3.0 & $ { 3.0 }^{ +0.0 }_{ -0.0 }$ & $0.95$ & $0.50$~$(0.00)$ \\
\RunThr SR1$\ell$    & 2 & 2.9$\pm$1.5 & 0.078 & 4.4 & $ { 5.0 }^{ +2.0 }_{ -1.2 }$ & $0.36$ & $0.50$~$(0.00)$ \\
\RunThr SR1$\ell$600 & 0 & 0.35$\pm$0.36 & 0.054 & 3.0 & $ { 3.0 }^{ +0.6 }_{ -0.0 }$ & $0.19$ & $0.50$~$(0.00)$ \\ \bottomrule
\end{tabular}
\end{center}

\end{table}

\subsection{Model-dependent exclusion limits}

As no significant excess is observed in the discovery regions, upper limits are set at 95\% CL for the simplified signal model for $\chinoonepm\chinoonemp$ and $\chinoonepm\ninoone$ production with RPV decays $\chinoonepm\rightarrow h\ell^\pm$ and $\ninoone\rightarrow h\nu$ described in Section~\ref{sec:intro}. Other decay modes of the charginos and neutralinos are assumed to be negligible after selections and are not considered. The limits are determined for each mass point through signal-plus-background fits using all CRs, \RunTwo and \RunThr SR1$\ell$, and \RunTwo and \RunThr SR2$\ell$, with each SR2$\ell$ binned in \mchinooneone. The \RunTwo and \RunThr counterparts of each region are treated as separate regions. The $\tilde{q}_\mu$ test statistic~\cite{Cowan:2010js} with the \CLs prescription~\cite{Read:2002hq} and the asymptotic approximation are used~\cite{Cowan:2010js}.

Results for the hypothesis where each flavor of charged lepton is equally likely are shown in Figure~\ref{fig:limits_lep}. SR2$\ell$ provides most of the sensitivity to low chargino and neutralino masses ($\lesssim\qty{500}{\gev}$), while SR1$\ell$ and SR2$\ell$ have similar sensitivity to high chargino and neutralino masses.
For the branching fractions $\mathcal{B}(\chinoonepm\rightarrow h\ell^\pm)=100\%$ and $\mathcal{B}(\ninoone\rightarrow h\nu)=100\%$, charginos and neutralinos with masses between $\qty{150}{\gev}$ and $\qty{1100}{\gev}$ are excluded. This has complementary sensitivity to the previous search for a trilepton resonance~\cite{SUSY-2018-36}, which excludes charginos and neutralinos with masses between $\qty{100}{\gev}$ and $\qty{975}{\gev}$ for the alternate $\mathcal{B}(\chinoonepm\rightarrow Z\ell^\pm)=100\%$ and $\mathcal{B}(\ninoone\rightarrow Z\nu)=100\%$ assumptions.

Results are also obtained for alternative flavor hypotheses. In addition to the usual SR requirements, the pure-electron (pure-muon) interpretation requires the reconstructed lepton in SR1$\ell$ and two leading reconstructed leptons in SR2$\ell$ to be electrons (muons). Figure~\ref{fig:limits_el} shows the upper limits for the hypothesis where the charged lepton from the decay is always an electron while Figure~\ref{fig:limits_mu} shows the upper limits for the hypothesis where the charged lepton from the decay is always a muon. Charginos and neutralinos with masses between $\qty{150}{\gev}$ and $\qty{1225}{\gev}$ ($\qty{150}{\gev}$ and $\qty{1150}{\gev}$) are excluded for the pure-electron (pure-muon) interpretation. Figure~\ref{fig:limits_tau} shows the upper limits for scenarios with $\mathcal{B}(\chinoonepm\rightarrow h\tau^\pm)=100\%$, excluding charginos and neutralinos with masses between $\qty{450}{\gev}$ and $\qty{750}{\gev}$. The pure-electron and pure-muon interpretations have greater sensitivity than the pure-$\tau$ scenario because the analysis does not reconstruct hadronically-decaying $\tau$-leptons. In addition, only part of the energy from leptonically-decaying $\tau$-leptons is retained by the daughter electron or muon, which can cause the lepton \pt to be below the \qty{40}{\GeV} threshold as well as lead to misreconstruction of the chargino pairs. This also causes the pure-electron and pure-muon scenarios to have higher sensitivity than the equal electron, muon, and $\tau$-lepton scenario. The pure-electron scenario has slightly higher sensitivity than the pure-muon scenario due to a higher electron reconstruction efficiency.

\begin{figure}[h!]
\centering
\includegraphics[width=0.49\linewidth]{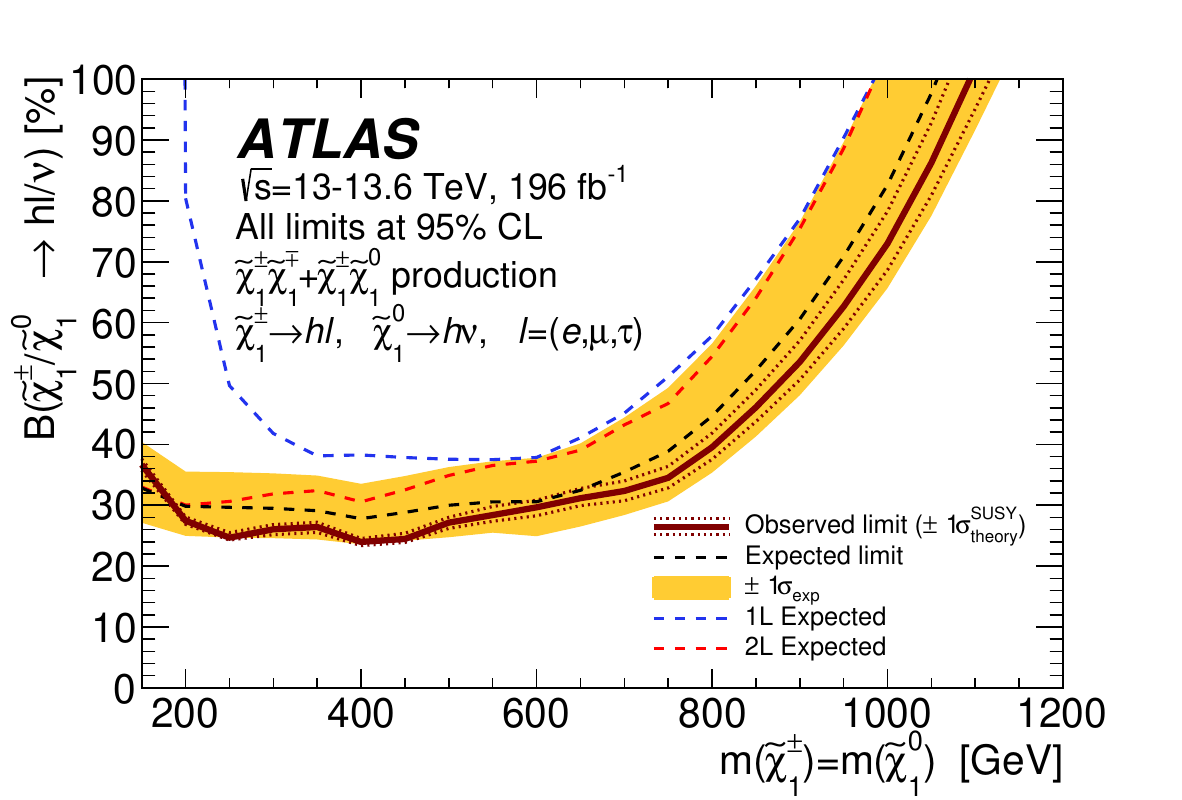}
\caption{95\% CL upper limits on $\mathcal{B}(\chinoonepm\rightarrow h\ell^\pm)$, with $\mathcal{B}(\ninoone\rightarrow h\nu)$ assumed to be equal, for the hypothesis where the charginos are equally likely to decay into each flavor of charged lepton. The expected and observed limits are shown by the dashed and solid lines respectively. The shaded bands around the expected limit show the $\pm1\sigma$ variations due to statistical and systematic uncertainties, while the dotted bands around the observed limit show the $\pm1\sigma$ variations from the theoretical uncertainty on the signal cross section. The expected limits using only 1-lepton (2-lepton) regions are shown by the dashed blue and dashed red lines respectively.}
\label{fig:limits_lep}
\end{figure}

\begin{figure}[h!]
\centering
\subfloat[]{
\centering
\includegraphics[width=0.49\linewidth]{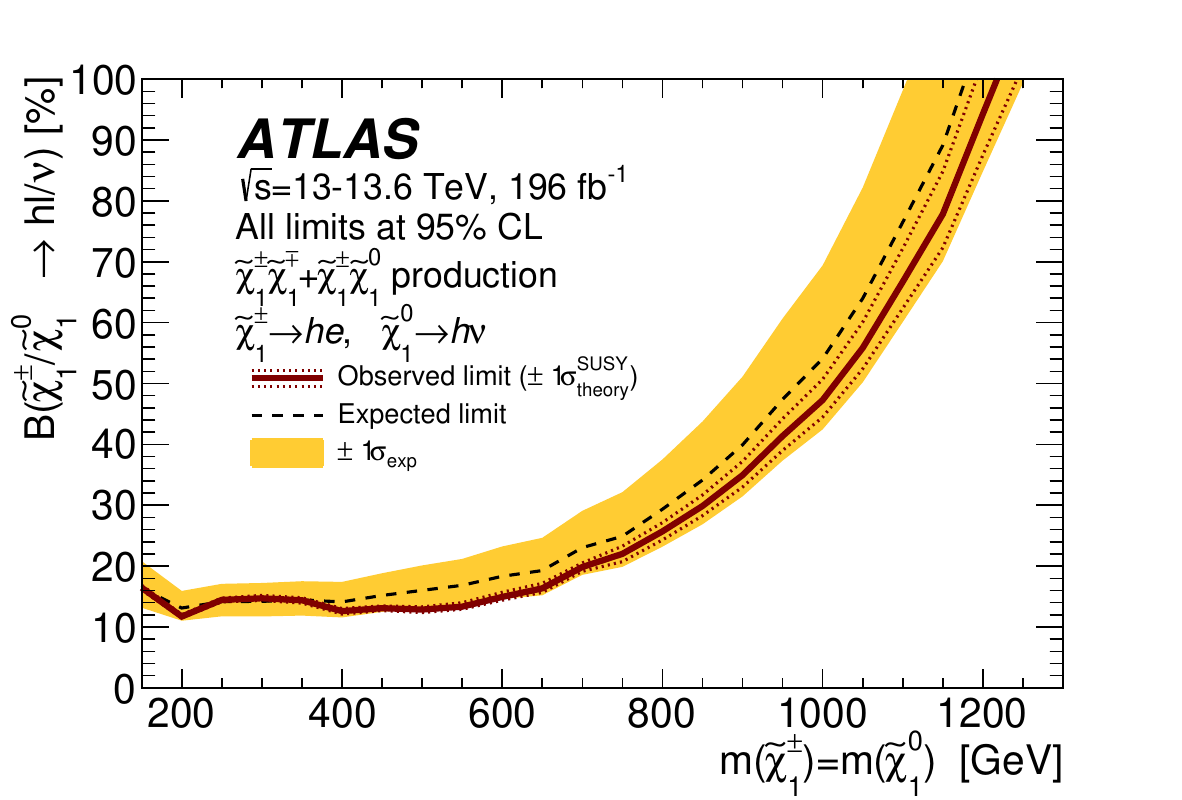}
\label{fig:limits_el}
}~
\subfloat[]{
\centering
\includegraphics[width=0.49\linewidth]{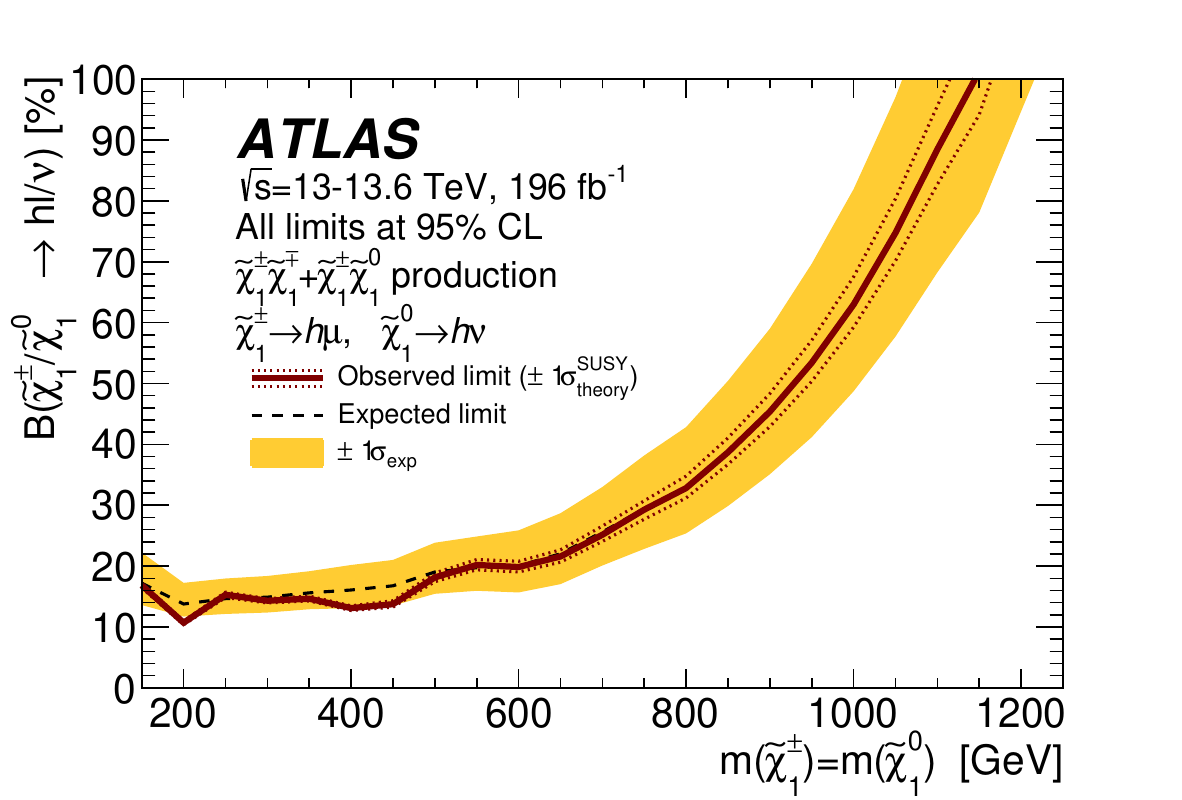}
\label{fig:limits_mu}
}
\caption{95\% CL upper limits on $\mathcal{B}(\chinoonepm\rightarrow h\ell^\pm)$, with $\mathcal{B}(\ninoone\rightarrow h\nu)$ assumed to be equal, for (a) the hypothesis where the charged lepton from the decay is always an electron and (b) the hypothesis where the charged lepton from the decay is always a muon. The expected and observed limits are shown by the dashed and solid lines respectively. The shaded bands around the expected limit show the $\pm1\sigma$ variations due to statistical and systematic uncertainties, while the dotted bands around the observed limit show the $\pm1\sigma$ variations from the theoretical uncertainty on the signal cross section.}
\label{fig:limits_elmu}
\end{figure}

\begin{figure}[h!]
\centering
\includegraphics[width=0.49\linewidth]{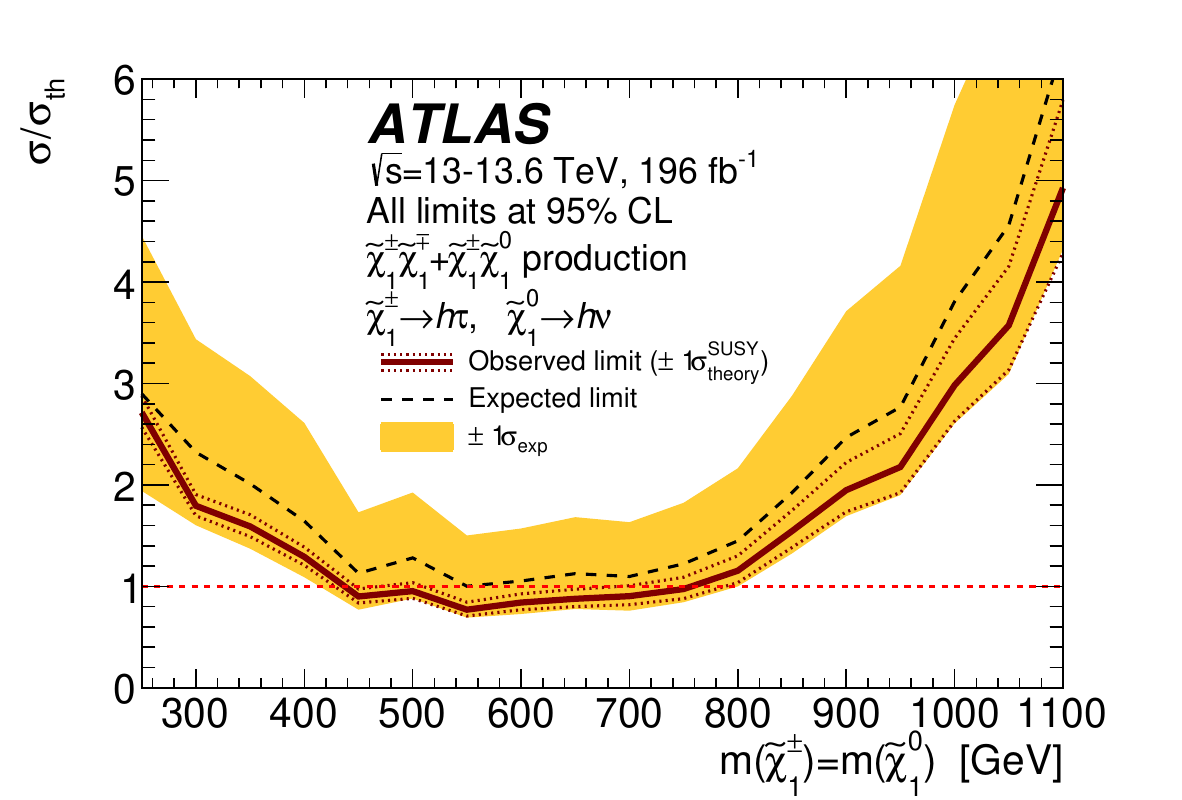}
\caption{95\% CL upper limits on the cross section for $\chinoonepm\chinoonemp$ and $\chinoonepm\ninoone$ production normalized to the theoretical prediction, assuming $\mathcal{B}(\chinoonepm\rightarrow h\tau^\pm)=100\%$ and $\mathcal{B}(\ninoone\rightarrow h\nu)=100\%$. The expected and observed limits are shown by the dashed and solid lines respectively. The shaded bands around the expected limit show the $\pm1\sigma$ variations due to statistical and systematic uncertainties, while the dotted bands around the observed limit show the $\pm1\sigma$ variations from the theoretical uncertainty on the signal cross section.}
\label{fig:limits_tau}
\end{figure}

% End of text imported from the .//sections/results.tex input file

%
%
%
%
%

%
\FloatBarrier

\section{Conclusion}
\label{sec:conclusion}

% The next lines are included from the .//sections/conclusion.tex input file
A search for $\chinoonepm\chinoonemp$ and $\chinoonepm\ninoone$ production where each chargino and neutralino decays into a Higgs boson and a lepton was performed in a final state with one or two charged leptons and at least four jets, with at least three identified $b$-jets. This search complements a previous search for RPV chargino decays into a $Z$ boson and a lepton in a channel with three or more leptons, and provides the first constraints on the simplified signal model for charginos with high decay branching fractions to Higgs bosons and leptons. The data correspond to an integrated luminosity of $\qty{196}{\ifb}$ of proton--proton collision data produced at center-of-mass energies of $\sqrt{s}=13$~\TeV and $\sqrt{s}=13.6$~\TeV, collected by the ATLAS experiment at the LHC between 2015 and 2023.

No significant excess above the SM prediction was observed. Model-independent limits are set at 95\% CL on the visible cross section for new physics processes. Charginos and neutralinos with masses between $\qty{150}{\gev}$ and $\qty{1100}{\gev}$ are excluded at 95\% CL for the $\mathcal{B}(\chinoonepm\rightarrow h\ell^\pm)=100\%$ and $\mathcal{B}(\ninoone\rightarrow h\nu)=100\%$ hypothesis, where each flavor of charged lepton (electron, muon, and $\tau$-lepton) is equally likely to be produced. Lower limits of $\qty{1225}{\gev}$ and $\qty{1150}{\gev}$ are also set on the chargino and neutralino masses for the $\mathcal{B}(\chinoonepm\rightarrow he^\pm)=100\%$ and $\mathcal{B}(\chinoonepm\rightarrow h\mu^\pm)=100\%$ hypotheses respectively.

% End of text imported from the .//sections/conclusion.tex input file

%
%
%
%
%
\section*{Acknowledgments}
%

%

% The next lines are included from the .//acknowledgements/Acknowledgements.tex input file
%
%

%
%

We thank CERN for the very successful operation of the LHC and its injectors, as well as the support staff at
CERN and at our institutions worldwide without whom ATLAS could not be operated efficiently.

The crucial computing support from all WLCG partners is acknowledged gratefully, in particular from CERN, the ATLAS Tier-1 facilities at TRIUMF/SFU (Canada), NDGF (Denmark, Norway, Sweden), CC-IN2P3 (France), KIT/GridKA (Germany), INFN-CNAF (Italy), NL-T1 (Netherlands), PIC (Spain), RAL (UK) and BNL (USA), the Tier-2 facilities worldwide and large non-WLCG resource providers. Major contributors of computing resources are listed in Ref.~\cite{ATL-SOFT-PUB-2026-001}.

We gratefully acknowledge the support of ANPCyT, Argentina; YerPhI, Armenia; ARC, Australia; BMWFW and FWF, Austria; ANAS, Azerbaijan; CNPq and FAPESP, Brazil; NSERC, NRC and CFI, Canada; CERN; ANID, Chile; CAS, MOST and NSFC, China; Minciencias, Colombia; MEYS CR, Czech Republic; DNRF and DNSRC, Denmark; IN2P3-CNRS and CEA-DRF/IRFU, France; SRNSFG, Georgia; BMFTR, HGF and MPG, Germany; GSRI, Greece; RGC and Hong Kong SAR, China; ICHEP and Academy of Sciences and Humanities, Israel; INFN, Italy; MEXT and JSPS, Japan; CNRST, Morocco; NWO, Netherlands; RCN, Norway; MNiSW, Poland; FCT, Portugal; MNE/IFA, Romania; MSTDI, Serbia; MSSR, Slovakia; ARIS and MVZI, Slovenia; DSI/NRF, South Africa; MICIU/AEI, Spain; SRC and Wallenberg Foundation, Sweden; SERI, SNSF and Cantons of Bern and Geneva, Switzerland; NSTC, Taipei; TENMAK, T\"urkiye; STFC/UKRI, United Kingdom; DOE and NSF, United States of America.

Individual groups and members have received support from BCKDF, CANARIE, CRC and DRAC, Canada; CERN-CZ, FORTE and PRIMUS, Czech Republic; COST, ERC, ERDF, Horizon 2020 and Marie Sk{\l}odowska-Curie Actions, European Union; Investissements d'Avenir Labex, Investissements d'Avenir Idex and ANR, France; DFG and AvH Foundation, Germany; Herakleitos, Thales and Aristeia programmes co-financed by EU-ESF and the Greek NSRF, Greece; BSF-NSF and MINERVA, Israel; NCN and NAWA, Poland; La Caixa Banking Foundation, CERCA and AGAUR programs from Generalitat de Catalunya and PROMETEO and GenT Programmes Generalitat Valenciana, Spain; G\"{o}ran Gustafssons Stiftelse, Sweden; The Royal Society and Leverhulme Trust, United Kingdom; Eric and Wendy Schmidt Fund for Strategic Innovation, United States of America.

In addition, individual members wish to acknowledge support from Chile: Agencia Nacional de Investigaci\'on y Desarrollo (ANID FONDECYT reg. 1230987, FONDECYT 1230812, FONDECYT 1240864, Fondecyt 3240661, Fondecyt Regular 1240721); China: Chinese Ministry of Science and Technology (MOST-2023YFA1605700, MOST-2023YFA1609300), National Natural Science Foundation of China (NSFC 12275265, NSFC-W2543005); Czech Republic: Czech Science Foundation (GACR - 24-11373S), Ministry of Education Youth and Sports (ERC-CZ-LL2327, FORTE CZ.02.01.01/00/22\_008/0004632), PRIMUS Research Programme (PRIMUS/21/SCI/017); EU: H2020 European Research Council (ERC - 101002463); European Union: European Research Council (BARD No. 101116429, ERC - 101219398, ERC - 948254, ERC 101089007), European Regional Development Fund (HE COFUND GA No.101081355, ERDF), Marie Sklodowska-Curie Actions (GAP-101168829); France: Agence Nationale de la Recherche (ANR-21-CE31-0013, ANR-22-EDIR-0002, ANR-24-CE31-0504-01); Germany: Deutsche Forschungsgemeinschaft (DFG - 469666862); China: Research Grants Council (GRF); Italy: Istituto Nazionale di Fisica Nucleare (LHC-MIUR - 28003/2025), Ministero dell'Università e della Ricerca (NextGenEU I53D23000820006 M4C2.1.1, SOE2024\_0000023); Japan: Japan Society for the Promotion of Science (JSPS KAKENHI  JP25H0063, JSPS KAKENHI JP22H01227, JSPS KAKENHI JP22H04944, JSPS KAKENHI JP22KK0227, JSPS KAKENHI JP24K23939, JSPS KAKENHI JP24KK0251, JSPS KAKENHI JP25H00650, JSPS KAKENHI JP25H01291, JSPS KAKENHI JP25K01011, JSPS KAKENHI JP25K01023); Poland: Polish National Science Centre (NCN 2021/42/E/ST2/00350, NCN OPUS 2023/51/B/ST2/02507, NCN OPUS nr 2022/47/B/ST2/03059, NCN UMO-2019/34/E/ST2/00393, UMO-2022/47/O/ST2/00148, UMO-2023/49/B/ST2/04085, UMO-2023/51/B/ST2/00920, UMO-2024/53/N/ST2/00869); Spain: Agència de Gestió d'Ajuts Universitaris i de Recerca. (AGAUR - 2023 BP 00141), Ministry of Science and Innovation (RYC2019-028510-I, RYC2020-030254-I, RYC2021-031273-I, RYC2022-038164-I), Ministerio de Ciencia, Innovación y Universidades/Agencia Estatal de Investigaci\'on (EU NextGenerationEU (PRTR-C17.I1), PID2022-142604OB-C22); Sweden: Carl Trygger Foundation (Carl Trygger Foundation CTS 22:2312), Swedish Research Council (Swedish Research Council 2023-04654, VR 2021-03651, VR 2022-03845, VR 2022-04683, VR 2023-03403, VR 2024-05451, VR 2025-05940), Knut and Alice Wallenberg Foundation (KAW 2023.0366); Switzerland: Swiss National Science Foundation (SNSF - PCEFP2\_194658); United Kingdom: The Binks Trust, Royal Society (NIF-R1-231091); United States of America: U.S. Department of Energy (ECA DE-AC02-76SF00515), John Templeton Foundation (John Templeton Foundation 63206), Neubauer Family Foundation.

%
%

% End of text imported from the .//acknowledgements/Acknowledgements.tex input file

%
\clearpage
\printbibliography
\clearpage
% ATLAS Collaboration author list
% Reference date of HMBS-2024-09 is 2026-03-31
% Author list last updated on date 25-JUN-26
% Data extracted on 25-Jun-2026 for paper reference HMBS-2024-09
% at 12:57am
 
\begin{flushleft}
\hypersetup{urlcolor=black}
{\Large The ATLAS Collaboration}

\bigskip

\AtlasOrcid[0000-0002-6665-4934]{G.~Aad}$^\textrm{\scriptsize 102}$,
\AtlasOrcid[0000-0001-7616-1554]{E.~Aakvaag}$^\textrm{\scriptsize 17}$,
\AtlasOrcid[0000-0002-5888-2734]{B.~Abbott}$^\textrm{\scriptsize 121}$,
\AtlasOrcid[0000-0002-0287-5869]{S.~Abdelhameed}$^\textrm{\scriptsize 83b}$,
\AtlasOrcid[0000-0002-1002-1652]{K.~Abeling}$^\textrm{\scriptsize 54}$,
\AtlasOrcid[0000-0001-5763-2760]{N.J.~Abicht}$^\textrm{\scriptsize 48}$,
\AtlasOrcid[0000-0002-8496-9294]{S.H.~Abidi}$^\textrm{\scriptsize 30}$,
\AtlasOrcid[0009-0003-6578-220X]{M.~Aboelela}$^\textrm{\scriptsize 44}$,
\AtlasOrcid[0000-0002-9987-2292]{A.~Aboulhorma}$^\textrm{\scriptsize 36e}$,
\AtlasOrcid[0000-0001-5329-6640]{H.~Abramowicz}$^\textrm{\scriptsize 154}$
% ...
\bigskip
\\

% ...
$^{17}$Department for Physics and Technology, University of Bergen, Bergen; Norway.\\
% ...
$^{30}$Physics Department, Brookhaven National Laboratory, Upton NY; United States of America.\\
% ...
$^{36}$$^{(a)}$Facult\'e des Sciences Ain Chock, Universit\'e Hassan II de Casablanca;$^{(b)}$Facult\'{e} des Sciences, Universit\'{e} Ibn-Tofail, K\'{e}nitra;$^{(c)}$Facult\'e des Sciences Semlalia, Universit\'e Cadi Ayyad, LPHEA-Marrakech;$^{(d)}$LPMR, Facult\'e des Sciences, Universit\'e Mohamed Premier, Oujda;$^{(e)}$Facult\'e des sciences, Universit\'e Mohammed V, Rabat;$^{(f)}$Institute of Applied Physics, Mohammed VI Polytechnic University, Ben Guerir; Morocco.\\
% ...
$^{44}$Physics Department, Southern Methodist University, Dallas TX; United States of America.\\
% ...
$^{48}$Fakult\"{a}t Physik, Technische Universit{\"a}t Dortmund, Dortmund; Germany.\\
% ...
$^{54}$II. Physikalisches Institut, Georg-August-Universit\"{a}t G\"ottingen, G\"ottingen; Germany.\\
% ...
$^{83}$$^{(a)}$Khalifa University of Science and Technology, Abu Dhabi;$^{(b)}$New York University Abu Dhabi, Abu Dhabi;$^{(c)}$United Arab Emirates University, Al Ain;$^{(d)}$University of Sharjah, Sharjah; United Arab Emirates.\\
% ...
$^{102}$CPPM, Aix-Marseille Universit\'e, CNRS/IN2P3, Marseille; France.\\
% ...
$^{121}$Homer L. Dodge Department of Physics and Astronomy, University of Oklahoma, Norman OK; United States of America.\\
% ...
$^{154}$Raymond and Beverly Sackler School of Physics and Astronomy, Tel Aviv University, Tel Aviv; Israel.\\
% ...
$^{a}$ Also at Affiliated with an institute formerly covered by a cooperation agreement with CERN.\\
$^{b}$ Also at An-Najah National University, Nablus; Palestine.\\
$^{c}$ Also at Borough of Manhattan Community College, City University of New York, New York NY; United States of America.\\
$^{d}$ Also at Center for Interdisciplinary Research and Innovation (CIRI-AUTH), Thessaloniki; Greece.\\
$^{e}$ Also at Centre of Physics of the Universities of Minho and Porto (CF-UM-UP); Portugal.\\
$^{f}$ Also at CERN, Geneva; Switzerland.\\
% ...

\end{flushleft}

% Created with Glance <Atlas.Glance@cern.ch>

%

%
%

%

%
%
%
%

\end{document}